\documentclass[10pt,journal,compsoc]{IEEEtran}

\ifCLASSOPTIONcompsoc
\else
\fi

\ifCLASSINFOpdf
\else
\fi

\usepackage{amsmath}
\usepackage{amssymb}
\usepackage{amsfonts}
\usepackage{amscd}
\usepackage{mathrsfs}
\usepackage{graphicx}
\usepackage[usenames,dvipsnames]{color}
\usepackage{bm}
\usepackage{url}
\usepackage{algorithm}
\usepackage{verbatim}
\usepackage{algpseudocode}

\newtheorem{definition}{Definition}
\newtheorem{proposition}{Proposition}

\newtheorem{example}{Example}

\newcommand{\AD}[1]{\textcolor{black}{{#1}}}

\newcommand{\FF}{\mathbb{F}}
\newcommand{\cv}{\mathbf{c}}
\newcommand{\xv}{\mathbf{x}}
\newcommand{\zv}{\mathbf{z}}

\newcommand{\Nc}{\mathcal{N}}

\usepackage{tikz}
\newcommand*\circled[1]{\tikz[baseline=(char.base)]{
            \node[shape=circle,draw,inner sep=0.5pt] (char) {#1};}}

\begin{document}
%
\title{Compressed Differential Erasure Codes\\ for Efficient Archival of Versioned Data}

\author{J. Harshan, Anwitaman Datta, Fr\'ed\'erique Oggier
\IEEEcompsocitemizethanks{\IEEEcompsocthanksitem J. Harshan and Fr\'ed\'erique Oggier are with the Division of Mathematical Sciences, Nanyang Technological University, Singapore. Anwitaman Datta is with the School of Computer Engineering, Nanyang Technological University, Singapore. Email:\{jharshan,anwitaman,frederique\}@ntu.edu.sg}
}

%
%

\markboth{March~2015}%
{Shell \MakeLowercase{\textit{et al.}}: Bare Advanced Demo of IEEEtran.cls for Journals}
%

\IEEEtitleabstractindextext{%
\begin{abstract}
In this paper, we study the problem of storing an archive of versioned data in a reliable and efficient manner in distributed storage systems. We propose a new storage technique called differential erasure coding (DEC) where the differences (deltas) between subsequent versions are stored rather than the whole objects, akin to a typical delta encoding technique. However, unlike delta encoding techniques, DEC opportunistically exploits the sparsity (i.e., when the differences between two successive versions have few non-zero entries) in the updates to store the deltas using compressed sensing techniques applied with erasure coding. We first show that DEC provides significant savings in the storage size for versioned data whenever the update patterns are characterized by in-place alterations. Subsequently, we propose a practical DEC framework so as to reap storage size benefits against not just in-place alterations but also real-world update patterns such as insertions and deletions \AD{that alter the overall data sizes}. We conduct experiments with several synthetic workloads to demonstrate that the practical variant of DEC provides significant reductions in storage overhead (up to 60\% depending on the workload) compared to baseline storage system which incorporates concepts from \emph{Rsync}, a delta encoding technique to store and synchronize data across a network. 
\end{abstract}

}

\maketitle


%

\vspace{-1cm}

%
%
\section{Introduction}
\label{sec1}

Distributed storage systems enable the storage of huge amount of data across networks of storage nodes. In such systems, redundancy of the stored data is critical to ensure fault tolerance against node failures. While data replication remains a practical way of realizing this redundancy, the past years have witnessed the adoption of erasure codes for data archival (e.g. in Microsoft Azure \cite{microsoft}, Hadoop FS \cite{hadoop}, or Google File System \cite{google}), which offer a better trade-off between storage overhead and fault tolerance. Design of erasure coding techniques amenable to reliable and efficient storage has accordingly garnered a huge attention \cite{DRWS,OgD2}.

Recent research works in distributed storage systems have predominantly focused on efficient storage of stand-alone data objects. Not many have addressed the aspect of efficiently storing multiple versions of data. Recently, Wang and Cadambe \cite{WaV} have addressed multi-version coding for distributed data, where the underlying problem is to encode different versions so that certain subsets of storage nodes can be accessed to retrieve the most common version among them. Their strategy has been shown applicable when the updates for the latest version do not reach all the nodes, possibly due to network problems. In this paper, we investigate a new aspect of erasure code design, aimed at storing multiple versions of data. The presented work is loosely related to the issues of efficient updates \cite{RVBS,HPZV,MWC,ECD}, and of deduplication \cite{dedup}. Existing works on update of erasure coded data focus on the computational and communication efficiency in carrying out the updates, with the goal to store only the latest version of the data, and thus do not delve into efficient storage or manipulation of the previous versions. Deduplication is the process of eliminating duplicate data blocks, which is used in order to eliminate unnecessary redundancy. Though we do not directly address it, the update semantic we use, that of storing only the differences across versions (akin to SVN \cite{svn}), means unnecessary duplicates are not created while storing the many versions, while our coding technique itself focuses on reducing the storage and I/O overheads of storing reliably the differences across versions.

The need to store multiple versions of data arise in many scenarios. For instance, when editing and updating files, users may want to explicitly create a version repository using a framework like SVN \cite{svn} or Git \cite{git}. Cloud based document editing or storage services also often provide the users access to older versions of the documents. Another scenario is that of system level back-up, where directories, whole file systems or databases are archived - and versions refer to the different system snapshots. In either of the two file centric settings, irrespective of whether a working copy used during editing is stored locally or on the cloud, or in a system level back up, say using copy-on-write \cite{copyonwrite}, the back-end storage system needs to preserve the different versions reliably, and can leverage on erasure coding for reducing the storage overheads. 

We propose a new differential erasure coding (DEC) framework that falls under the umbrella of delta encoding techniques, where the differences (deltas) between subsequent versions are stored rather than the whole objects. The proposed technique exploits the sparsity in the differences among versions by applying techniques from compressed sensing \cite{compsens}, in order to reduce the storage overhead (see Sections \ref{sec2} and \ref{sec3}). We have already proposed the idea of combining compressed sensing with erasure coding in \cite{HOD}, where we studied the benefits purely in terms of I/O, and for objects of fixed size. While retaining the combination of compressed sensing and erasure coding, this work introduces a different erasure coding strategy which provides storage overhead benefits. We first present a simplistic layout of DEC that relies on fixed object lengths across successive versions of the data, so as to evaluate the right choice of erasure codes to store versioned data. We show that when all the versions are fetched in ensemble, there is also an equivalent gain in I/O operations. This comes at an increased I/O overhead when accessing individual versions. We accordingly propose some heuristics to optimize the basic DEC, and demonstrate that they ameliorate the drawbacks adequately without compromising the gains at all. Further, we show that the combination of compressed sensing and erasure coding yields other practical benefits such as the possibility of employing fewer erasure codes against different sparsity levels of the update patterns (see Section \ref{sec4}).  

In the later part of this paper, we extend the preliminary ideas of DEC to develop a framework for practical DEC that is robust to real-world update patterns across versions such as insertions and deletions \AD{which may alter the overall size of the data object}. Along that direction, we acknowledge that insertions and deletions may ripple changes across the object at the coding granularity, and may also increase the object size. Such rippling effect could in particular render DEC useless, and obliterate the consequent benefits. To circumvent such hurdles, we apply zero padding schemes, an obvious way to ameliorate the aforementioned problems, taking into account insertions, deletions and in-place alterations (see Section \ref{sec:pdec}). For storing versioned data, the total storage size for \emph{deltas} inclusive of zero pads (prior to erasure coding) is used as the metric to evaluate the quantum and placement of zero pads against a wide range of workloads that include insertions and deletions, both bursty and distributed in nature (see Section \ref{sec7}). \AD{As a baseline, we choose an intuitive setup} that uses concepts from Rsync \cite{rsync}, a delta encoding technique to store and synchronize files over the network, to store successive versions of the object. We compare the storage savings offered by the practical DEC technique with the baseline and show that the savings from DEC can soar as high as 60\% depending on the distribution of the workload.

\subsection{System Model for Version Management}

\begin{figure}
\includegraphics[scale=0.55]{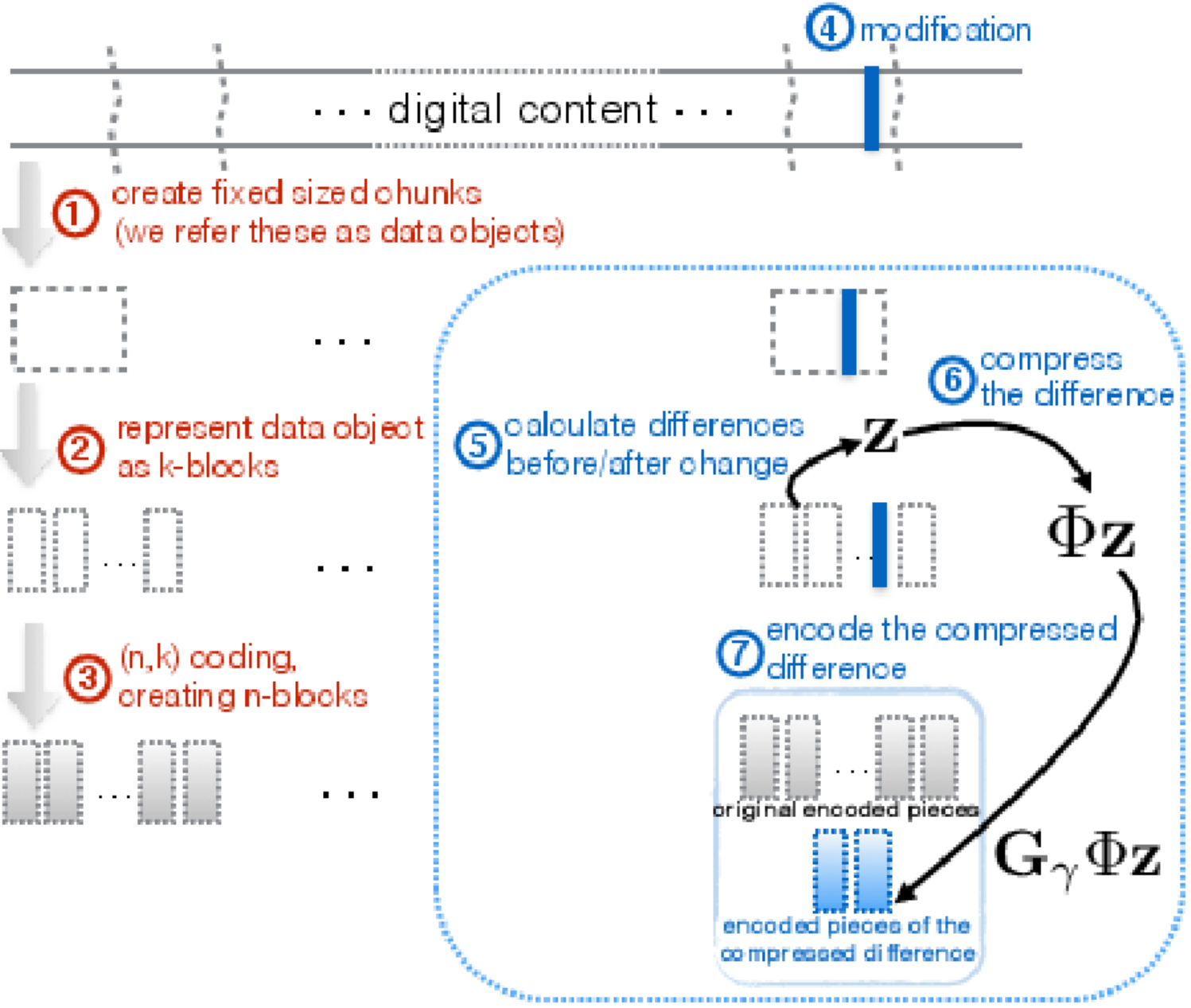}
\caption{An overview of the coding strategy using compressed differences
\label{fig:overview}
}
\end{figure}

Any digital content to be stored, be it a file, directory, database, or a whole file system, is divided into data chunks, shown as phase \textcolor{red}{\circled{1}} in Fig. \ref{fig:overview}. The proposed coding techniques are agnostic of the nuances of the upper levels, and all subsequent discussions will be at the granularity of these chunks, which we will refer to as data objects or just objects. 

Formally, we denote by $\xv \in \FF_q^k$ a data object to be stored over a network, that is, the object is seen as a vector of $k$ blocks (phase \textcolor{red}{\circled{2}}) taking value in the alphabet $\FF_q$, with $\FF_q$ the finite field with $q$ elements, $q$ a power of 2 typically.
Encoding for archival of an object $\xv$ across $n$ nodes is done (phase \textcolor{red}{\circled{3}}) using an $(n,k)$ linear code, that is $\xv$ is mapped to the codeword
\begin{equation}\label{eq:lincode}
\cv = \mathbf{G}\xv \in \FF_q^n,~n>k,
\end{equation}
for $\mathbf{G}$ an $n\times k$ generator matrix with coefficients in $\FF_q$.
We use the term {\em systematic} to refer to a codeword $\cv$ whose $k$ first components are $\xv$, 
that is $c_i=x_i$, $i=1,\ldots,k$. This described what is a standard encoding procedure used in erasure coding based storage systems. We suppose next that the content mutates, and we wish to store all the versions.

Let $\xv_1 \in \FF_q^k$ denote the first version of a data object to be stored. When it is modified (phase \textcolor{blue}{\circled{4}}), a new version $\xv_2 \in \FF_q^k$ of this object is created. More generally, a new version $\mathbf{x}_{j+1}$ is obtained from $\xv_j$ to produce over time a sequence $\{ \mathbf{x}_{j} \in \mathbb{F}^{k}_{q},~j=1,2,\ldots, L < \infty\}$ of $L$ different versions of a data object, to be stored in the network. 
We are not concerned with the application level semantic of the modifications, but with the bit level changes in the object.
Thus the changes between two successive versions are captured by the relation
\begin{equation}
\label{update_eq}
\mathbf{x}_{j + 1} = \mathbf{x}_{j} + \mathbf{z}_{j + 1},
\end{equation}
where $\mathbf{z}_{j + 1} \in \mathbb{F}^{k}_{q}$ denotes the modifications (in phase \textcolor{blue}{\circled{5}}) of the $j$th update. We first assume fixed object lengths across successive versions of data so as to build an uncomplicated framework for the differential strategy. Such a framework shields us from unnecessarily delving into system specificities, instead, serves as a foundation to evaluate various erasure coding techniques to store multiple versions of data. We show that the design, analysis and assessment of the coding techniques are oblivious to the nuances of how the data object is broken down into several chunks prior to the encoding purposes, thereby facilitating us to segregate \emph{chunk synthesis} and \emph{erasure coding} blocks as two independent entities. In the later part of this work (see Section \ref{sec:pdec}), we discuss how to relax the fixed object length assumption and yet develop a practical DEC scheme, that is robust to variable object lengths across successive versions.

The key idea is that when the changes from $\xv_{j}$ to $\xv_{j+1}$ are small (decided by the sparsity of $\zv_{j+1}$), it is possible to apply compressed sensing \cite{compsens}, which permits to represent a $k$-length $\gamma$-sparse vector $\mathbf{z}$ (see Definition \ref{def:gamma}) with less than $k$ components (phase \textcolor{blue}{\circled{6}}) through a linear transformation on $\mathbf{z}$, which does not depend on the position of the non-zero entries, in order to gain in  storage efficiency.

\begin{definition}\label{def:gamma}
For some integer $1 \leq \gamma < k$, a vector $\zv \in \mathbb{F}^{k}_{q}$ is said to be $\gamma$-sparse if it contains at most $\gamma$ non-zero entries. 
\end{definition}

Let $\zv \in \mathbb{F}^{k}_{q}$ be $\gamma$-sparse such that $\gamma < \frac{k}{2}$, and $\Phi \in \mathbb{F}^{2\gamma \times k}_{q}$ denote the {\em measurement matrix} used for compressed sensing. The compressed representation $\zv' \in \mathbb{F}^{2\gamma}_{q}$ of $\zv$ is obtained as 
\begin{equation}\label{eq:compsens}
\zv' = \Phi \zv.
\end{equation}

The following proposition \cite{ZP} gives a sufficient condition on $\Phi$ to uniquely recover $\zv$ from $\zv'$ using a syndrome decoder.  
\begin{proposition}
\label{prop1}
If any $2\gamma$ columns of $\Phi$ are linearly independent, the $\gamma$-sparse vector $\zv$ can be recovered from $\zv'$.
\end{proposition}

Once sparse modifications are compressed, which reduces the I/O reads, they are encoded into codewords of length $<n$  (phase \textcolor{blue}{\circled{7}}) decreasing in turn the storage overhead.

%
%
%
\section{Differential Erasure Encoding for Version-control}
\label{sec2}

Let $\{\mathbf{x}_{j} \in \mathbb{F}^{k}_{q},~1\leq j \leq L\}$ be the sequence of versions of a data object to be stored. The changes from $\mathbf{x}_{j}$ to $\mathbf{x}_{j + 1}$ are reflected in the vector $\mathbf{z}_{j + 1}=\xv_{j+1}-\xv_j$ in \eqref{update_eq} which is $\gamma_{j+1}$-sparse (see Definition \ref{def:gamma}) for some $1 \leq \gamma_{j+1} \leq k$. The value $\gamma_{j+1}$ may a priori vary across versions of one object, and across application domains.  
All the versions $\xv_1,\ldots,\xv_L$ need protection from node failures, and are archived using a linear erasure code (see (\ref{eq:lincode})).

\subsection{Object Encoding}
\label{sec2_subsec1}

We describe a generic {\em differential} encoding (called {\bf Step $j+1$}) suited for efficient archival of versioned data, which exploits the sparsity of the updates, when $\gamma_{j+1} < \tfrac{k}{2}$, to reduce the storage overheads of archiving all the versions reliably.
We assume that one storage node is in possession of two versions, say $\xv_j$ and $\xv_{j+1}$ of one data object, $j=1,\ldots,L-1$. The corresponding implementation is discussed in Subsection \ref{sec2_subsec:impl}.  

{\bf Step $j+1$.} Two versions $\xv_j$ and $\xv_{j+1}$ are located in one storage node. 
The difference vector $\mathbf{z}_{j+1} = \mathbf{x}_{j+1} - \mathbf{x}_{j}$ and the corresponding sparsity level $\gamma_{j+1}$ are computed. 
If $\gamma_{j+1} \geq \frac{k}{2}$, the object $\mathbf{z}_{j+1}$ is encoded as $\mathbf{c}_{j+1} = \mathbf{G}\mathbf{z}_{j+1}$. On the other hand, if $\gamma_{j+1} < \frac{k}{2}$, then $\mathbf{z}_{j+1}$ is first compressed (see (\ref{eq:compsens})) as 
\begin{equation*}
\mathbf{z}'_{j+1} = \Phi_{\gamma_{j+1}}\mathbf{z}_{j+1}, 
\end{equation*}
where $\Phi_{\gamma_{j+1}} \in \mathbb{F}^{2\gamma_{j+1} \times k}_{q}$ is a measurement matrix such that any $2\gamma_{j+1}$ of its columns are linearly independent (see Proposition \ref{prop1}). Subsequently, $\mathbf{z}'_{j+1}$ is encoded as 
\[
\mathbf{c}_{j+1} = \mathbf{G}_{\gamma_{j+1}}\mathbf{z}'_{j+1}, 
\]
where $\mathbf{G}_{\gamma_{j+1}} \in \mathbb{F}^{n_{\gamma_{j+1}} \times 2\gamma_{j+1}}_{q}$ is the generator matrix of an $(n_{\gamma_{j+1}}, 2\gamma_{j+1})$ erasure code with storage overhead $\kappa$. The components of $\mathbf{c}_{j+1}$ are distributed across a set $\mathcal{N}_{j+1}$ of $n_{\gamma_{j+1}}$ nodes, whose choice is discussed in Section 
\ref{sec2_subsec:impl}.

Since $\gamma_{j+1}$ is random, a total of $\lceil \frac{k}{2} \rceil $ erasure codes denoted by $\mathcal{G} = \{\mathbf{G}, \mathbf{G}_{1}, \ldots, \mathbf{G}_{\lceil \frac{k}{2} \rceil - 1} \}$, and a total of $\lceil \frac{k}{2} \rceil - 1$ measurement matrices denoted by $\Sigma = \{\Phi_{1}, \Phi_{2}, \ldots, \Phi_{\lceil \frac{k}{2} \rceil - 1} \}$ have to be designed a priori. The erasure codes may be taken systematic and/or MDS (that is, such that any $n-k$ failure patterns are tolerated), our scheme works irrespective of these choices. 
This encoding strategy implies one extra matrix multiplication whenever a sparse difference vector is obtained.

We give a toy example to illustrate the computations.
\begin{example}\label{ex:ex1}
Take $k=4$, suppose that the digital content is written in binary as $(100110010010)$ and that the linear code used for storage is a $(6,4)$ code over $\FF_8$. To create the first data object $\xv_1$, cut the digital content into $k=4$ chunks $100$, $110$, $010$, $010$, so that $\xv_1$ is written over $\FF_8$ as $\xv_1=(1,1+w,w,w)$ where $w$ is the generator of $\FF^*_8$, satisfying $w^3=w+1$. The next version of the digital content is created, say $(10011011001)$. Similarly $\xv_2$ becomes $\xv_2=(1,1+w,1+w,w)$, and the difference vector $\zv_2$ is given by $\zv_2=\xv_2-\xv_1=(0,0,1,0)$, with $\gamma_2=1 < k/2$. Apply a measurement matrix $\Phi_{\gamma_2}=\Phi_1$ to compress $\zv_2$:
\[
\Phi_1\zv_2=
\begin{bmatrix}
1 & 0 & w & w+1 \\
0 & 1 & w+1 & w
\end{bmatrix}
\begin{bmatrix}
0 \\
0 \\
1 \\
0
\end{bmatrix}
=
\begin{bmatrix}
w \\
w+1 
\end{bmatrix}=\zv_2'.
\]  
Note that every two columns of $\Phi_1$ are linearly independent (see Proposition \ref{prop1}), thus allowing the compressed vector to be recovered.
Encode $\zv_2'$ using a single parity check code:
\[
\cv_2=
\begin{bmatrix}
1 & 0 \\
0 & 1 \\
1 & 1
\end{bmatrix}
\begin{bmatrix}
w \\
w+1
\end{bmatrix}
=
\begin{bmatrix}
w \\
w+1 \\
1
\end{bmatrix}.
\]
\end{example}

\begin{figure}
\begin{algorithmic}[1]
\Procedure{Encode}{$\mathcal{X}, \mathcal{G}, \Sigma$}
\State \textbf{FOR} $0 \leq j \leq L-1$
\State $~~~$ \textbf{IF} $j = 0$
\State $~~~$ $~~~$ return $\mathbf{c}_{1} = \mathbf{G}\mathbf{x}_{1}$;
\State $~~~$ \textbf{ELSE} (\emph{This part summarizes Step $j+1$ in the text})
\State $~~~$ $~~~$ Compute $\mathbf{z}_{j+1} = \mathbf{x}_{j+1} - \mathbf{x}_{j}$;
\State $~~~$ $~~~$ Compute $\gamma_{j+1}$;
\State $~~~$ $~~~$ \textbf{IF} $\gamma_{j+1} \geq \frac{k}{2}$
\State $~~~$ $~~~$ $~~~$ return $\mathbf{c}_{j+1} = \mathbf{G}\mathbf{z}_{j+1}$;
\State $~~~$ $~~~$ \textbf{ELSE}
\State $~~~$ $~~~$ $~~~$ Compress $\mathbf{z}_{j+1}$ as $\mathbf{z}'_{j+1} = \Phi_{\gamma_{j+1}}\mathbf{z}_{j+1}$;
\State $~~~$ $~~~$ $~~~$ return $\mathbf{c}_{j+1} = \mathbf{G}_{\gamma_{j+1}}\mathbf{z}_{j+1}$;
\State $~~~$ $~~~$ \textbf{END IF}
\State $~~~$\textbf{END IF}
\State \textbf{END FOR}
\EndProcedure
\end{algorithmic}
\caption{Encoding Procedure for DEC}\label{Algorithm1}
\end{figure}

\subsection{Implementation and Placement}
\label{sec2_subsec:impl}

{\bf Caching.} To store $\xv_{j+1}$ for $j \geq 1$, the proposed scheme requires the calculation of differences between the existing version $\xv_j$ and the new version $\xv_{j+1}$ in (\ref{update_eq}). However, it does not store $\xv_j$, but $\xv_1$ together with $\zv_2,\ldots,\zv_j$. Reconstructing $\xv_j$ before computing the difference and encoding the new difference is expensive in terms of I/O operations, network bandwidth, latency as well as computations. A practical remedy is thus to cache a full copy of the latest version $\xv_j$, until a new version $\xv_{j+1}$ arrives. This also helps in improving the response time and overheads of data read operations in general, and thus disentangles the system performance from the storage efficient resilient storage of all the versions.  

Considering caching as a practical method, an algorithm summarizes the differential erasure coding (DEC) procedure in Fig. \ref{Algorithm1}. The input and the output of the algorithm are $\mathcal{X} = \{\mathbf{x}_{j} \in \mathbb{F}^{k}_{q},~1\leq j \leq L\}$ and $\{\mathbf{c}_{j},~1\leq j \leq L\}$, respectively.  

{\bf Placement consideration.} The choice of the sets $\Nc_{j+1}$, $j=0,\ldots,L-1$  of nodes over which the different versions are stored needs a closer introspection. Since $\xv_1$ together with $\zv_2,\ldots,\zv_j$ are needed to recover $\xv_j$ (see also Subsection \ref{subsec:retrDEC}), if $\xv_1$ is lost, $\xv_j$ cannot be recovered, and thus there is no gain in fault tolerance by storing $\xv_j$ in a different set of nodes than $\Nc_1$. Furthermore, since $n_{\gamma_j}<n$, codewords $\cv_i$s may have different resilience to failures. The dependency of $\xv_j$ on previous versions suggests that the fault-tolerance of subsequent versions are determined by the worst fault-tolerance achieved among $\cv_i$s for $i<j$.

\begin{example}
We continue Example \ref{ex:ex1}, where $\xv_1$ is encoded into $\cv_1=(c_{11},\ldots,c_{16})$ using a $(6, 4)$ MDS code. Allocate $c_{1i}$ to $N_i$, that is use the set $\Nc_1=\{N_1,\ldots,N_6\}$ of nodes. Store $\cv_2$ in $\Nc_2=\{N_1, N_2, N_3\} \subset \mathcal{N}_1$ for collocated placement, and in $\Nc_2=\{N'_1, N'_2, N'_3\}$, $\Nc_2 \cap \Nc_1 = \emptyset$ for distributed placement. Let $p$ be the probability that a node fails, and failures are assumed independent. We compute the probability to recover both $\xv_1$ and $\xv_2$ in case of node failures (known as {\em static resilience}) for both distributed and collocated strategies.

\begin{figure}
\begin{center}
\includegraphics[width=3in]{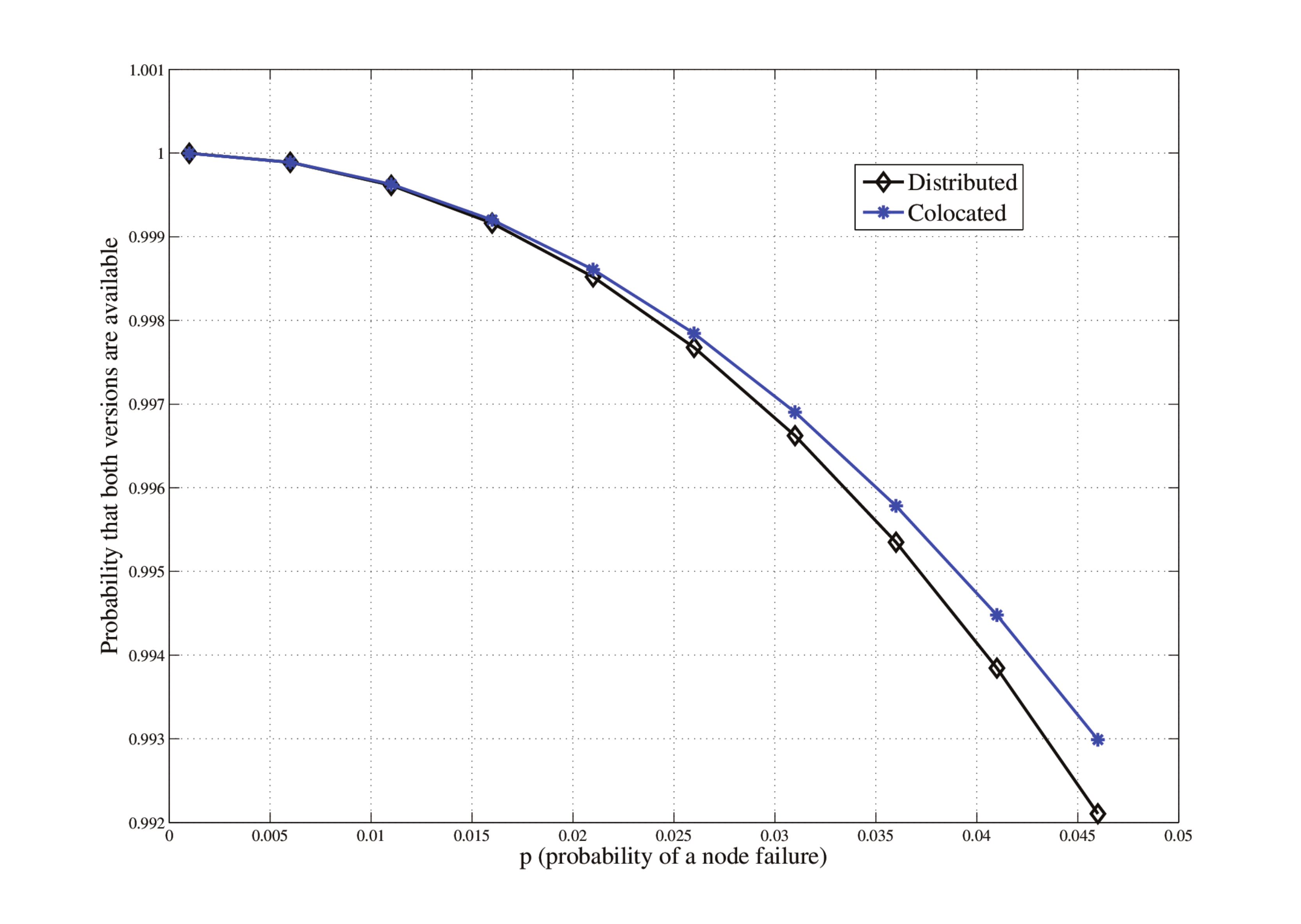}
\vspace{-0.4cm}
\caption{Placement consideration: comparing probability that both versions are available}
\label{fig_placement_example}
\end{center}
\end{figure}
\end{example}

For distributed placement, the set of error events for losing $\mathbf{x}_{1}$ is $\mathcal{E}_{1} = \{\mbox{3 or more nodes fail in } \mathcal{N}_{1}\}.$ Hence, the probability $\mbox{Prob}(\mathcal{E}_{1})$ of losing $\textbf{x}_{1}$ is given by
\begin{equation}
\label{dr1}
p^6 + C^{6}_{5} p^5(1-p) + C^{6}_{4} p^4(1-p)^{2}  + C^{6}_{3} p^3(1-p)^{3},
\end{equation}
where $C^{m}_{r}$ denotes the $m$ choose $r$ operation. The set of error events for losing $\mathbf{z}_{2}$ stored with a (3,2) MDS code is $\mathcal{E}_{2} = \{\mbox{2 or 3 nodes fail in } \mathcal{N}_{2}\}.$ Thus, $\textbf{z}_{2}$ is lost with probability 
\begin{eqnarray}
\label{dr2}
\mbox{Prob}(\mathcal{E}_{2}) & = & p^3 + C^{3}_{2} p^2(1-p).
\end{eqnarray}
From \eqref{dr1} and \eqref{dr2}, the probability of retaining both versions is
\begin{eqnarray}
\label{final1}
\mbox{Prob}_{d}(\mathbf{x}_{1}, \mathbf{x}_{2}) \triangleq (1- \mbox{Prob}(\mathcal{E}_{1}))(1- \mbox{Prob}(\mathcal{E}_{2})).
\end{eqnarray}
The set of error events for losing $\mathbf{x}_{1}$ or $\mathbf{z}_{2}$ is $$\mathcal{E}_{1} \cup \mathcal{E}_{2} = \{\mbox{3 or more nodes fail}\} \cup \{\mbox{specific 2 nodes failure}\}$$
for collocated placement.
 Out of $C^{6}_{2}$ possible 2 node failure patterns, 3 patterns contribute to the loss of the object $\mathbf{z}_{2}$. Therefore, $\mbox{Prob}(\mathcal{E}_{1} \cup \mathcal{E}_{2})$ is
\begin{equation*}
p^6 + C^{6}_{5} p^5(1-p) + C^{6}_{4} p^4(1-p)^{2} + 
C^{6}_{3} p^3(1-p)^{3} + 3p^2(1-p)^{4}
\end{equation*}
from which, the probability of retaining both the versions is
\begin{eqnarray}
\label{final2}
\mbox{Prob}_{c}(\mathbf{x}_{1}, \mathbf{x}_{2}) \triangleq 1- \mbox{Prob}(\mathcal{E}_{1} \cup \mathcal{E}_{2}).
\end{eqnarray}
In Fig. \ref{fig_placement_example}, we compare \eqref{final1} and \eqref{final2} for different values of $p$ from $0.001$ to $0.05$. The plot shows that collocated allocation results in better resilience than the distributed case. 

{\bf Optimized Step $j+1$.} Based on these insights, a practical change of {\bf Step $j$} is:
if $\gamma_{j+1} \geq \frac{k}{2}$, $\zv_{j+1}$ is discarded and $\xv_{j+1}$ is encoded as $\mathbf{c}_{j+1} = \mathbf{G}\mathbf{x}_{j+1}$, to ensure that a whole version is again encoded. Since many contiguous sparse versions may be created, we put as a heuristic an iteration threshold $\iota$, after which even if all differences from one version to another stay very sparse, a whole version is used for coding and storage.


\begin{center}
\begin{table*}
\caption{I/O access metrics for the traditional and the differential schemes to store $\{\mathbf{x}_{1}, \mathbf{x}_{2}, \ldots, \mathbf{x}_{l}\}$}
\begin{center}
\begin{tabular}{|c|c|c|c|c|c|c|c|c|c|c|}
\hline Parameter & Traditional & Forward Differential & Reverse differential\\
\hline I/O reads to read the $l$-th version & $k$ & $k + \sum_{j = 2}^{l} \mbox{min} (2 \gamma_{j}, k)$ & $k$\\
\hline I/O reads to read the first $l$-th versions & $lk$ & $k + \sum_{j = 2}^{l} \mbox{min} (2 \gamma_{j}, k)$ & $k + \sum_{j = 2}^{l} \mbox{min} (2 \gamma_{j}, k)$\\
\hline Number of Encoding Operations & 1 (on the latest version) & 1 (on the latest version) & 2 (on the latest and the preceding version)\\
\hline Total Storage Size till the $l$-th version & $ln$ & $n + \sum_{j = 2}^{l} \mbox{min}(2 \kappa \gamma_{j}, n)$ & $n + \sum_{j = 2}^{l} \mbox{min}(2 \kappa \gamma_{j}, n)$\\
\hline
\end{tabular}
\end{center}
\label{table1}
\end{table*}
\end{center}

\subsection{On the Storage Overhead}

Since employed erasure codes depend on the sparsity level, the storage overhead of the above differential encoding improves upon that of encoding different versions independently. The average gains in storage overhead are discussed in Section \ref{sec:simulation}. Formally, the total storage size till the $l$-th version is 
\begin{equation*}
\delta(\mathbf{x}_{1}, \mathbf{x}_{2}, \ldots, \mathbf{x}_{l}) = n + \sum_{j = 2}^{l} \min(2 \kappa \gamma_{j}, n) \leq ln,
\end{equation*}
for $2 \leq l \leq L$. The storage overhead for the {\bf Optimized Step $j+1$} is the same as that of {\bf Step $j+1$} since for $\gamma_{j+1} \geq \frac{k}{2}$, the coded objects $\mathbf{G}\mathbf{x}_{j+1}$ and $\mathbf{G}\mathbf{z}_{j+1}$ have the same size.

\subsection{Object Retrieval}
\label{subsec:retrDEC}

Suppose that $L$ versions of a data object are archived using {\bf Step $j+1$}, $j \leq L-1$ and the user needs to retrieve some $\mathbf{x}_{l}$, $1 < l \leq L$. Assuming that there are enough encoded blocks for each $\mathbf{c}_i$ ($i\leq l$) available, relevant nodes in the sets $\mathcal{N}_{1}, \ldots, \mathcal{N}_{l}$ are accessed to fetch and decode the $\cv_i$ to obtain $\mathbf{x}_{1}$, and the $l-1$ compressed differences $\mathbf{z}'_{2}, \mathbf{z}'_{3}, \ldots, \mathbf{z}'_{l}$. See Subsection \ref{sec2_subsec:impl} for a discussion on placement and an illustration that reusing the same set of nodes gives the best availability with MDS codes, hence bounding the number of accessed nodes by $|\mathcal{N}_{1}|$. All compressed differences sharing the same sparsity can be added first, and then decompressed, since 
\[
\sum_{i \in J_{\gamma}} \mathbf{z}'_{i}= \Phi_{\gamma} \sum_{i \in J_{\gamma}} \mathbf{z}_{i}
\]
for $J_{\gamma}=\{ j|\gamma_j=\gamma\}$.
The cost of recovering $\sum_{i \in J_{\gamma}} \mathbf{z}_{i} $ is only one decompression instead of $|J_{\gamma}|$,
with which $\mathbf{x}_{l}$ is given by
\begin{equation*}
\mathbf{x}_{l} = \mathbf{x}_{1} + \sum_{j = 2}^{l} \mathbf{z}_{j}.
\end{equation*}

A minimum of $k$ I/O reads is needed to retrieve $\mathbf{x}_{1}$. For $\mathbf{z}_{j}$ ($2 \leq j \leq l$), the number of I/O reads may be lower than $k$, depending on the update sparsity. 
If $\gamma_{j} < \frac{k}{2}$, then $\mathbf{z}'_{j}$ is retrieved with $2\gamma_{j}$ I/O reads, while if $\gamma_{j} \geq \frac{k}{2}$, then $\mathbf{z}_{j}$ is recovered with $k$ I/O reads, so that $\mbox{min} (2 \gamma_{j}, k)$ I/O reads are needed for $\mathbf{z}_{j}$. The total number of I/O reads to retrieve $\mathbf{x}_{l}$  is 
\begin{equation}
\label{reads_dec}
\eta(\mathbf{x}_{l}) = k + \sum_{j = 2}^{l} \mbox{min} (2 \gamma_{j}, k)
\end{equation}
and so is the total number of I/O reads to retrieve the first $l$ versions: $\eta(\mathbf{x}_{1}, \mathbf{x}_{2}, \ldots, \mathbf{x}_{l}) = \eta(\mathbf{x}_{l})$.

To retrieve $\mathbf{x}_{l}$ for $1 \leq l \leq L$, when archival was done using {\bf Optimized Step $j+1$}, $j\leq L-1$, look for the most recent version $\mathbf{x}_{l'}$ such that  $l' \leq l$ and $\gamma_{l'} \geq \frac{k}{2}$. Then, using $\{\mathbf{x}_{l'}, \mathbf{z}_{l'+1}, \ldots, \mathbf{z}_{l} \}$, the object $\mathbf{x}_{l}$ is reconstructed as $\mathbf{x}_{l} = \mathbf{x}_{l'} + \sum_{j = l'+1}^{l} \mathbf{z}_{j}.$ Hence, the total number of I/O reads is 
\begin{equation}
\label{reads_dec_opt}
\eta(\mathbf{x}_{l}) = k + \sum_{j = l' + 1}^{l} \mbox{min} (2 \gamma_{j}, k).
\end{equation}
The number of I/O reads to retrieve the first $l$ versions is the same as for {\bf Step $j+1$}.

The benefits of the proposed differential encoding in terms of average number of I/O reads are presented in Section \ref{sec:simulation}.
 
\begin{example}\label{ex:L20}
Assume that $L = 20$ versions of an object of size $k = 10$ are differentially encoded, with sparsity profile $\{\gamma_{j},~ 2 \leq j \leq L\}$ $=\{3,$$ 8,$$ 3,$$ 6,$$ 7,$$ 9,$$ 10,$$ 6,$$ 2,$$ 2,$$ 3, $$9, $$3,$$ 9,$$ 3, $$10,$$ 4,$$ 2, $$3\}$. The storage pattern is $\{\mathbf{x}_{1}, $$\mathbf{z}_{2},$$ \mathbf{z}_{3},$$ \ldots,$$ \mathbf{z}_{20}\}$. Assuming $\mathbf{x}_{1}$ is not sparse, the  I/O read numbers to access $\{\mathbf{x}_{1},$$ \mathbf{z}_{2},$$ \mathbf{z}_{3},$$ \ldots,$$ \mathbf{z}_{20}\}$ are $\{10, 6, $$10, $$6,  10,  $$ 10,  10,  $$10,  10,$$ 4,  4,  6, $$ 10,  6,  10, $$ 6,  10,  8, $$ 4,  6\}$. The total I/O reads to recover all the 20 versions is 156 (instead of 200 for the non-differential method). The total storage space for all the 20 versions assuming a storage overhead of 2 is 312 (instead of 400 otherwise). The I/O read numbers to recover $\{\mathbf{x}_{1}, $$\mathbf{x}_{2}, \mathbf{x}_{3}, \ldots,$$ \mathbf{x}_{20}\}$ are $\{10, $$ 16, $$ 26, $$ 32, $$ 42, $$ 52,  $$62, $$ 72, $$ 82, $$ 86, $$ 90,$$ 96,$$ 106,$$ 112,$$ 122, $$128,$$ 138,\\ $$146,$$ 150, $$156\}$, while for the optimized step, we get $\{10, $$ 16,$$ 10, $\\$16, $$ 10, $$ 10,  $$10, $$ 10,$$  10, $$ 14, $$ 18,$$  24, $$ 10, $$ 16, $$ 10, $$ 16, $$ 10, $$ 18,$$ 22, $$ 28\}$.
\end{example}

%
%
%

\section{Reverse Differential Erasure Coding}
\label{sec3}

In Table \ref{table1}, we summarize the total storage size and the number of I/O reads required by the (forward) differential method. If some $\gamma_{j}$, $1 \leq j \leq l$, are smaller than $\frac{k}{2}$, then the number of I/O reads for joint retrieval of all the versions $\{\mathbf{x}_{1}, \mathbf{x}_{2}, \ldots, \mathbf{x}_{l} \}$ is lower than that of the traditional method. However, this advantage comes at the cost of higher number of I/O reads for accessing the $l$-th version $\mathbf{x}_{l}$ alone. Therefore, 
for applications where the latest archived versions are more frequently accessed than the joint versions, the overhead for reading the latest version dominates the advantage of reading multiple versions. For such applications, we propose a variant of the differential method called the reverse DEC, wherein the order of storing the difference vectors is reversed.

\subsection{Object Encoding}
\label{sec3_subsec1}

As in Subsection \ref{sec2_subsec1}, we assume that one node stores the latest version $\xv_j$ and the new version $\xv_{j+1}$ of a data object. Since $\xv_j$ is readily obtained, caching is less critical here.

{\bf Step $j+1$.} Compute the difference vector $\mathbf{z}_{j+1} = \mathbf{x}_{j+1} - \mathbf{x}_{j}$ and its sparsity level $\gamma_{j+1}$. The object $\mathbf{x}_{j+1}$ is encoded as $\mathbf{c}_{j+1} = \mathbf{G}\mathbf{x}_{j+1}$ and stored in $\mathcal{N}_{j+1}$. Furthermore, if $\gamma_{j+1} < \frac{k}{2}$, then $\mathbf{z}_{j+1}$ is first compressed as $\mathbf{z}'_{j+1} = \Phi_{\gamma_{j+1}}\mathbf{z}_{j+1},$ and then encoded as 
$
\mathbf{c} = \mathbf{G}_{\gamma_{j+1}}\mathbf{z}'_{j+1}, 
$
where $\mathbf{G}_{\gamma_{j+1}}$ is the generator matrix of an $(n_{\gamma_{j+1}}, 2\gamma_{j+1})$ erasure code. Finally, the preceding version $\mathbf{c}_{j}$ is overwritten as $\mathbf{c}_{j} = \mathbf{c}$. 

A key feature is that in addition to encoding the latest version $\mathbf{x}_{j+1}$, the preceding version is also re-encoded depending on the sparsity level $\gamma_{j+1}$, resulting in two encoding operations (instead of one for the method in Subsection \ref{sec2_subsec1}).

A summary of the encoding is provided in Fig. \ref{Algorithm2}. The storage overhead for this method is the same as the one in Section \ref{sec2}. The considerations on data placement and static resilience of $\cv_j$ in the set $\Nc_j$ of nodes are analogous as well, and an optimized version is obtained similarly as for the forward differential encoding. 

\begin{figure}
\begin{algorithmic}[1]
\Procedure{Encode}{$\mathcal{X}, \mathcal{G}, \Sigma$}
\State \textbf{FOR} $0 \leq j \leq L-1$
\State $~~~$ \textbf{IF} $j = 0$
\State $~~~$ $~~~$ return $\mathbf{c}_{1} = \mathbf{G}\mathbf{x}_{1}$;
\State $~~~$ \textbf{ELSE} (\emph{This part summarizes Step $j+1$ in the text})
\State $~~~$ $~~~$ $\mathbf{c}_{j+1} = \mathbf{G}\mathbf{x}_{j+1}$;
\State $~~~$ $~~~$ Compute $\mathbf{z}_{j+1} = \mathbf{x}_{j+1} - \mathbf{x}_{j}$;
\State $~~~$ $~~~$ Compute $\gamma_{j+1}$;
\State $~~~$ $~~~$ \textbf{IF} $\gamma_{j+1} < \frac{k}{2}$
\State $~~~$ $~~~$ $~~~$ Compress $\mathbf{z}_{j+1}$ as $\mathbf{z}'_{j+1} = \Phi_{\gamma_{j+1}}\mathbf{z}_{j+1}$;
\State $~~~$ $~~~$ $~~~$ return $\mathbf{c}_{j} = \mathbf{G}_{\gamma_{j+1}}\mathbf{z}_{j+1}$;
\State $~~~$ $~~~$ \textbf{END IF}
\State $~~~$\textbf{END IF}
\State \textbf{END FOR}
\EndProcedure
\end{algorithmic}
\caption{Encoding Procedure for the Reverse DEC}\label{Algorithm2}
\end{figure}

\subsection{Object Retrieval}

Suppose that $l$ versions of a data object have been archived, and the user needs to retrieve the latest version $\mathbf{x}_{l}$. In the reverse DEC, unlike Subsection \ref{sec2_subsec1}, the latest version $\mathbf{x}_{l}$ is encoded as $\mathbf{G}\mathbf{x}_{l}$. Hence, the user must access a minimum of $k$ nodes from the set $\mathcal{N}_{l}$ to recover $\mathbf{x}_{l}$. To retrieve all the $l$ versions $\{\mathbf{x}_{1}, \mathbf{x}_{2}, \ldots, \mathbf{x}_{l} \}$, the user accesses the nodes in the sets $\mathcal{N}_{1}, \mathcal{N}_{2}, \ldots, \mathcal{N}_{l}$ to retrieve $\mathbf{z}'_{2}, \mathbf{z}'_{3}, \ldots, \mathbf{z}'_{l}, \mathbf{x}_{l}$, respectively. The objects $\mathbf{z}_{2}, \mathbf{z}_{3}, \ldots, \mathbf{z}_{l}$ are recovered from $\mathbf{z}'_{2}, \mathbf{z}'_{3}, \ldots, \mathbf{z}'_{l}$, respectively through a sparse-reconstruction procedure, and $\mathbf{x}_{j}$, $1 \leq j \leq l-1$, is recursively reconstructed as 
\begin{equation*}
\mathbf{x}_{j} = \mathbf{x}_{l} - \left(\sum_{t = j}^{l} \mathbf{z}_{t}\right).
\end{equation*}
It is clear that a total of $k + \sum_{j = 2}^{l} \mbox{min} (2 \gamma_{j}, k)$ reads are needed for accessing all the $l$ versions and only $k$ reads for the latest version. The performance metrics of the reverse DEC scheme are also summarized in Table \ref{table1} (the last column).

\begin{example}
\label{ex:L20:reverse}
For the sparsity profile of Example \ref{ex:L20}, the storage pattern using reverse DEC is $\{\mathbf{z}_{2}, \mathbf{z}_{3}, \ldots, \mathbf{z}_{20}, \mathbf{x}_{20}\}$. The I/O read numbers to access $\{\mathbf{z}_{2}, \mathbf{z}_{3}, \ldots, \mathbf{z}_{20}, \mathbf{x}_{20}\}$ are $\{6,10,  6, 10, 10, 10, 10, 10, 4, 4, 6, 10, 6, 10, 6, 10, 8, 4, 6, 10\}$. The total storage size and the I/O reads to recover all the 20 versions are the same as that of the forward differential method. The I/O numbers to recover $\{\mathbf{x}_{1}, \mathbf{x}_{2}, \mathbf{x}_{3}, \ldots, \mathbf{x}_{20}\}$ are $\{156, 150, 144, 134, 124, 114, 104, 94, 84, 80, 76, 70, 60, 54, 44,$\\$ 38, 28,  20, 16, 10\}$. Note that I/O number to access the latest version (in this case 20th version) is lower than that of the forward differential scheme. For the optimized step, the corresponding I/O numbers are  $\{16, 10,$$ 16, 10, 10, 10,  10, 10,$\\$ 24, 20, 16, 10, 16, 10, 16, 10, 28, 20, 16, 10\}$.
\end{example}

\section{Two-level Differential Erasure Coding}
\label{sec4}

The differential encoding (both forward and the reverse DEC) exploits the sparse nature of the updates to reduce the storage size and the number of I/O reads. Such advantages stem from the application of $\lceil\frac{k}{2}\rceil$ erasure codes matching the different levels of sparsity ($\lceil\frac{k}{2}\rceil - 1$ erasure codes for each $\gamma < \frac{k}{2}$ and one for $\gamma \geq \frac{k}{2}$). If $k$ is large, then the system needs a large number of erasure codes, resulting in an impractical strategy. In this section, we employ only two erasure codes, termed {\em two-level differential erasure coding}, for the sake of easier implementation, and refer to the earlier differential schemes in Sections \ref{sec2}-\ref{sec3} as $\frac{k}{2}$-level DEC schemes. We need the following ingredients for the two-level DEC scheme:

(1) An $(n, k)$ erasure code with generator matrix $\mathbf{G} \in \mathbb{F}^{n \times k}_{q}$ to store the original data object. 

(2) A measurement matrix $\Phi_{T} \in \mathbb{F}^{2T \times k}_{q}$ to compress sparse updates, where $T \in \{ 1, 2, \ldots, \lfloor\frac{k}{2}\rfloor \}$ is a chosen  threshold. 

(3) An $(n_{T}, 2T)$ erasure code with generator matrix $\mathbf{G}_{T} \in \mathbb{F}^{n_{T} \times 2T}_{q}$ to store the compressed data object. The number $n_{T}$ is chosen such that $\kappa \triangleq \frac{n}{k} = \frac{n_{T}}{2T}$.

We discuss only the two-level \emph{forward} DEC scheme. The two-level \emph{reverse} DEC scheme is a straightforward variation.

\subsection{Object Encoding}

The key point of this encoding is that the number of erasure codes (and the corresponding measurement matrices) to store the $\gamma$-sparse vectors for $1 \leq \gamma < \frac{k}{2}$ is reduced from $\lceil \frac{k}{2} \rceil - 1$ to $1$. Thus, based on the sparsity level, the update vector is either compressed and then archived, or archived as it. Formally:

{\bf Step $j+1$.} Once the version $\mathbf{x}_{j+1}$ is created, using $\mathbf{x}_{j}$ in the cache, the difference vector $\mathbf{z}_{j+1} = \mathbf{x}_{j+1} - \mathbf{x}_{j}$ and the corresponding sparsity level $\gamma_{j+1}$ are computed. If $\gamma_{j+1} > T$, the object $\mathbf{z}_{j+1}$ is encoded as $\mathbf{c}_{j+1} = \mathbf{G}\mathbf{z}_{j+1}$, else $\mathbf{z}_{j+1}$ is first compressed (see (\ref{eq:compsens})) as 
$
\mathbf{z}'_{j+1} = \Phi_{T}\mathbf{z}_{j+1}, 
$
where the measurement matrix $\Phi_{T} \in \mathbb{F}^{2T \times k}_{q}$ is such that any $2T$ of its columns are linearly independent (see Proposition \ref{prop1}). Then, $\mathbf{z}'_{j+1}$ is encoded as 
$
\mathbf{c}_{j+1} = \mathbf{G}_{T}\mathbf{z}'_{j+1}, 
$
where $\mathbf{G}_{T} \in \mathbb{F}^{n_{T} \times 2T}_{q}$ is the generator matrix of an $(n_{T}, 2T)$ erasure code.  The components of $\mathbf{c}_{j+1}$ are stored across the set $\mathcal{N}_{j+1}$ of nodes.

A summary of the encoding method is provided in Fig. \eqref{Algorithm3}.

\begin{figure}
\begin{algorithmic}[1]
\Procedure{Encode}{$\mathcal{X}, \mathbf{G}, \mathbf{G}_{T}, \Phi_{T}$}
\State \textbf{FOR} $0 \leq j \leq L-1$
\State $~~~$ \textbf{IF} $j = 0$
\State $~~~$ $~~~$ return $\mathbf{c}_{1} = \mathbf{G}\mathbf{x}_{1}$;
\State $~~~$ \textbf{ELSE} (\emph{This part summarizes Step $j+1$ in the text})
\State $~~~$ $~~~$ Compute $\mathbf{z}_{j+1} = \mathbf{x}_{j+1} - \mathbf{x}_{j}$;
\State $~~~$ $~~~$ Compute $\gamma_{j+1}$;
\State $~~~$ $~~~$ \textbf{IF} $\gamma_{j+1} > T$
\State $~~~$ $~~~$ $~~~$ return $\mathbf{c}_{j+1} = \mathbf{G}\mathbf{z}_{j+1}$;
\State $~~~$ $~~~$ \textbf{ELSE}
\State $~~~$ $~~~$ $~~~$ Compress $\mathbf{z}_{j+1}$ as $\mathbf{z}'_{j+1} = \Phi_{T}\mathbf{z}_{j+1}$;
\State $~~~$ $~~~$ $~~~$ return $\mathbf{c}_{j+1} = \mathbf{G}_{T}\mathbf{z}_{j+1}$;
\State $~~~$ $~~~$ \textbf{END IF}
\State $~~~$\textbf{END IF}
\State \textbf{END FOR}
\EndProcedure
\end{algorithmic}
\caption{Encoding Procedure for Two-level DEC}\label{Algorithm3}
\end{figure}

\subsection{On the Storage Overhead}

The total storage size for the two-level DEC is $\delta(\mathbf{x}_{1}, \mathbf{x}_{2}, \ldots, \mathbf{x}_{l}) = n + \sum_{j = 2}^{l} n_{j},$ where
\begin{eqnarray}
\label{storage_two_level}
n_{j} = \left\{ \begin{array}{ccccc}
n, \mbox{ if } \gamma_{j} > T \\
\kappa 2T, \mbox{otherwise}.
\end{array} 
\right.
\end{eqnarray}

\subsection{Data Retrieval}

Similarly to the $\frac{k}{2}$-level DEC scheme, the object $\mathbf{x}_{l}$ for some $1 \leq l \leq L$ is reconstructed as 
$\mathbf{x}_{l} = \mathbf{x}_{1} + \sum_{j = 2}^{l} \mathbf{z}_{j},$ 
by accessing the nodes in the sets $\mathcal{N}_{1}, \mathcal{N}_{2}, \ldots, \mathcal{N}_{l}$. To retrieve $\mathbf{x}_{1}$, a minimum of $k$ I/O reads is needed. If $\mathbf{z}_{j}$ is $\gamma_{j}$-sparse and $\gamma_{j} \leq T$, then $\mathbf{z}'_{j}$ is first retrieved with $2T$ I/O reads, second, $\mathbf{z}_{j}$ is decoded from $\mathbf{z}'_{j}$ and $\Phi_{T}$ through a sparse-reconstruction procedure. On the other hand, if $\gamma_{j} > T$, then $\mathbf{z}_{j}$ is recovered with $k$ I/O reads. Overall, the total number of I/O reads for $\mathbf{x}_{l}$ in the differential set up is 
$
\eta(\mathbf{x}_{l}) = k + \sum_{j = 2}^{l} \eta_{j},
$
where 
\begin{eqnarray}
\label{read_two_level}
\eta_{j} = \left\{ \begin{array}{ccccc}
2T, \mbox{ if } \gamma_{j} \leq T \\
k, \mbox{otherwise}.
\end{array} 
\right.
\end{eqnarray}
Similarly, the total number of I/O reads to retrieve the first $l$ versions is also $\eta(\xv_1,\ldots,\xv_l) = k + \sum_{j = 2}^{l} \eta_{j}$.

\begin{example}\label{ex:2level}
We apply the threshold $T = 3$ to the sparsity profile in Example \ref{ex:L20}. The object $\mathbf{z}_{18}$ (with $\gamma_{18} = 4$) is then archived without compression whereas all objects with sparsity lower than or equal to 3 are compressed using a $6 \times 10$ measurement matrix. The  I/O read numbers to access $\{\mathbf{x}_{1},$$ \mathbf{z}_{2},$$ \mathbf{z}_{3},$$ \ldots,$$ \mathbf{z}_{20}\}$ are $\{10, 6, $$10, $$6,  10,  $$ 10,  10,  $$10,  10,$$ 6,  6,  6, $$ 10,  6,  10, $$ 6,  10,  10, $$ 6,  6\}$.  The total number of I/O reads to access all the versions is $164$ and the corresponding storage size is $328$. Thus, with just two levels of compression, the storage overhead is more than the $5$-level DEC scheme but still lower than $400$.  
\end{example}

\subsection{Threshold Design Problem}
\label{sec4_subsec4}

For the two-level DEC, the total number of I/O reads and the storage size are random variables that are respectively given by
$\eta = k + \sum_{j = 2}^{L} \eta_{j},$ where $\eta_{j}$ is given in \eqref{read_two_level} and $\delta = n + \sum_{j = 2}^{L} n_{j},$ where $n_{j}$ is given in \eqref{storage_two_level}. Note that $\eta$ and $\delta$ are also dependent on the threshold $T$. The threshold $T$ that minimizes the average values of $\eta$ and $\delta$ is given by:
\begin{equation}
\label{th_prob}
T_{\small{\mbox{opt}}} = \arg \min_{T\in \{1, 2, \ldots, \lfloor \frac{k}{2} \rfloor \}} w \mathbb{E}[\delta(\mathbf{x}_{1}, \mathbf{x}_{2})] + (1-w)\mathbb{E}[\eta(\mathbf{x}_{1}, \mathbf{x}_{2})],
\end{equation}
where $0 \leq w \leq 1$ is a parameter that appropriately weighs the importance of storage overhead and I/O reads overhead, and $\mathbb{E}[\cdot]$ is the expectation operator over the random variables $\{ \Gamma_{2}, \Gamma_{3}, \ldots, \Gamma_{L}\}$. This optimization depends on the underlying probability mass functions (PMFs) on $\{\Gamma_{j}\}$, so we discuss the choice of the parameter $1 \leq T \leq \lfloor \frac{k}{2} \rfloor$ in Section \ref{sec:simulation}.

\subsection{Cauchy Matrices for Two-level DEC} 

Suppose that $\Phi_{T} \in \mathbb{F}^{2T \times k}_{q}$ is carved from a Cauchy matrix \cite{McS}.
A Cauchy matrix is such that any square submatrix is full rank \cite{LaF}. Thus, there exists a $2\gamma_{j} \times k$ submatrix $\Phi_{T}(\mathcal{I}_{2\gamma_{j}},:)$ of $\Phi_{T}$, where $\mathcal{I}_{2\gamma_{j}} \subset \{1, 2, \ldots, 2T\}$ represents the indices of $2\gamma_{j}$ rows, for which any $2\gamma_{j}$ columns are linearly independent, implying that the observations $\mathbf{r} = \Phi_{T}(\mathcal{I}_{2\gamma_{j}},:)\mathbf{z}_{j},$ can be retrieved from $\mathcal{N}_{j}$ with $2\gamma_{j}$ I/O reads. Also, using $\mathbf{r}$ and $\Phi_{T}(\mathcal{I}_{2\gamma_{j}},:)$, the sparse update $\mathbf{z}_{j}$ can be decoded through a sparse-reconstruction procedure. Thus, the number of I/O reads to get $\mathbf{z}_{j}$ is reduced from $2T$ to $2\gamma_{j}$ when $\gamma_{j} \leq T$. This procedure is applicable for any $\gamma_{j} < T$. Therefore, a $\gamma_{j}$-sparse vector with $\gamma_{j} \leq T$ can be recovered with $2\gamma_{j}$ I/O reads. The total number of I/O reads for $\mathbf{x}_{l}$ in the two-level DEC with Cauchy matrix is finally
$
\eta(\mathbf{x}_{l}) = k + \sum_{j = 2}^{l} \eta_{j},
$
where 
\begin{eqnarray}
\label{read_two_level_T}
\eta_{j} = \left\{ \begin{array}{ccccc}
2\gamma_{j}, \mbox{ if } \gamma_{j} \leq T \\
k, \mbox{otherwise}.
\end{array} 
\right.
\end{eqnarray}

Since the number of I/O reads is potentially different compared to the case without Cauchy matrices, the threshold design problem in $\eqref{th_prob}$ can result in different answers for this case. We discuss this optimization problem in Section \ref{sec:simulation}.

\begin{example}
With Cauchy matrix for $\Phi_{T}$ in Example \ref{ex:2level}, the I/O numbers to access $\{\mathbf{z}_{2}, \mathbf{z}_{3}, \ldots, \mathbf{z}_{20}, \mathbf{x}_{20}\}$ are $\{10, 6, $$10, $$6,  10,  $$ 10,  10,  $$10,  10,$$ 4,  4,  6, $$ 10,  6,  10, $$ 6,  10,  10, $$ 4,  6\}$, which makes the total I/O reads 158. However, the total storage size with Cauchy matrix continues to be $328$.
\end{example}

\begin{figure}
\begin{center}
\includegraphics[width=3.7in]{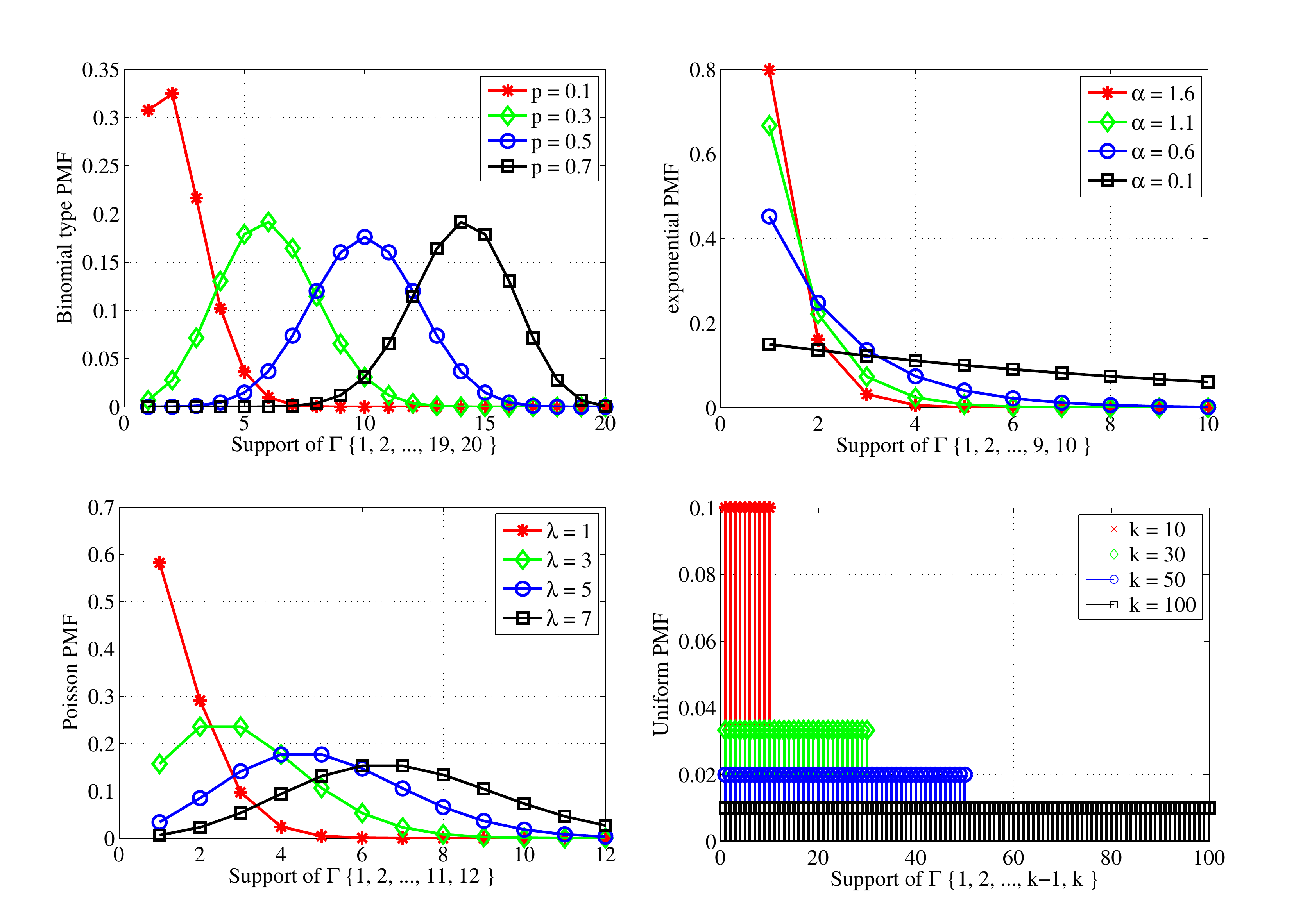}
\vspace{-0.5cm}
\caption{From top left, clock-wise: Binomial type PMF in $p$ (for k = 20), Truncated exponential PMF in $\alpha$ (for k = 10), Truncated Poisson PMF in $\lambda$ (for k = 12) and the uniform PMF for different object lengths $k$. The x-axis of these plots represent the support $\{1, 2, \ldots, k \}$ of the random variable $\Gamma$.}
\label{fig_1}
\end{center}
\end{figure}

%
%

\section{Simulation Results} 
\label{sec:simulation}

In this section, we present experimental results on the storage size and the number of I/O reads for the different differential encoding schemes. We assume that $\{\Gamma_{j},~ 2 \leq j \leq L\}$ is a set of random variables and its realizations $\{\gamma_{j},~ 2 \leq j \leq L\}$ are known. First we consider a version-control system with $L = 2$, which is the worst-case choice of $L$ as more versions could reveal more storage savings. This setting both (i) serves as a proof of concept, and (ii) already shows the storage savings for this simple case. Later, we also present experimental results for a setup with $L > 2$ versions.

%
%

\subsection{System with $L = 2$ versions}
\label{subsec:pmf}

For $L=2$, there is one random variable denoted henceforth as $\Gamma$, with realization $\gamma$. Since $\Gamma$ is a discrete random variable with finite support, we test the following finite support distributions for our experimental results on the average number of I/O reads for the two versions and the average storage size. 

\textbf{Binomial type PMF:} This is a variation of the standard Binomial distribution given by
\begin{equation}
\label{normalised_binomial}
\mbox{P}_{\Gamma}(\gamma) = c\frac{k!}{\gamma! (k-\gamma)!} p^{\gamma} (1-p)^{k-\gamma},~\gamma = 1, 2, \ldots, k,
\end{equation}
where $c = \frac{1}{1 - (1-p)^k}$ is the normalizing constant. The change is necessary since $\gamma =0$ is not a valid event. 

\textbf{Truncated exponential PMF:} This is a finite support version of the exponential distribution in parameter $\alpha > 0$:
\begin{equation}
\label{normalised_exponential}
\mbox{P}_{\Gamma}(\gamma) = c e^{-\alpha \gamma}.
\end{equation}
The constant $c$ is chosen such that $\sum_{\gamma = 1}^{k} \mbox{P}_{\Gamma}(\gamma) = 1$. 

\textbf{Truncated Poisson PMF:} This is a finite support version of the Poisson distribution in parameter $\lambda$ given by
\begin{equation}
\label{normalised_poisson}
\mbox{P}_{\Gamma}(\gamma) = c \frac{\lambda^{\gamma} e^{-\lambda}}{\gamma!},
\end{equation}
where the constant $c$ is chosen such that $\sum_{\gamma = 1}^{k} \mbox{P}_{\Gamma}(\gamma) = 1$ 

\textbf{Uniform PMF:} This is the standard uniform distribution: 
\begin{equation}
\label{uniform_dist}
\mbox{P}_{\Gamma}(\gamma) = \frac{1}{k}.
\end{equation}

\begin{figure}
\begin{center}
\includegraphics[width=3.8in]{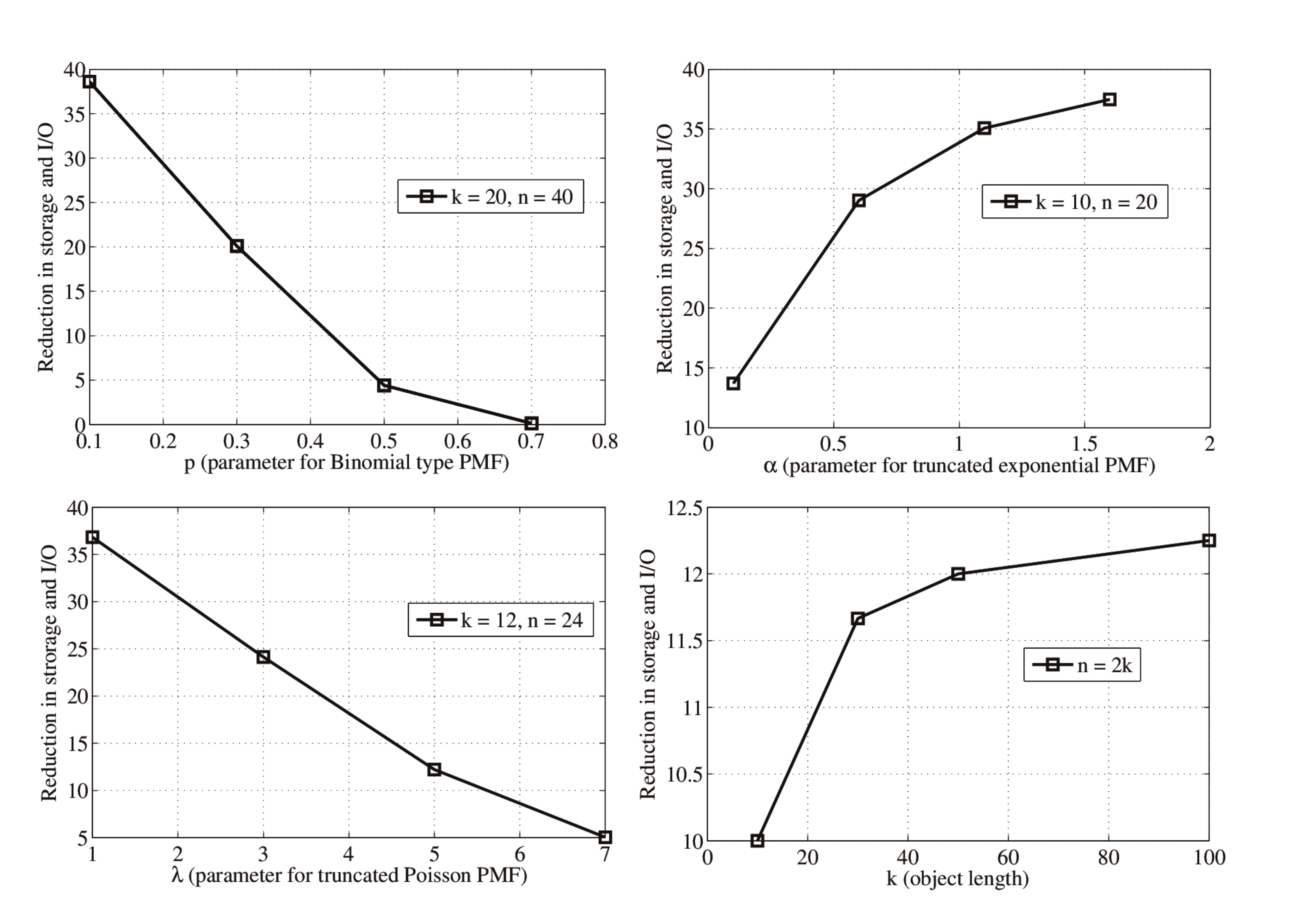}
\vspace{-0.7cm}
\caption{Average percentage reduction in the I/O reads and storage size for PMFs in Fig. \ref{fig_1} when $L = 2$. The experimental results are presented in the same order as that of the PMFs in Fig. \ref{fig_1}.}
\label{fig_2}
\end{center}
\end{figure}

In Fig. \ref{fig_1}, we plot the PMFs in \eqref{normalised_binomial}, \eqref{normalised_exponential}, \eqref{normalised_poisson} and \eqref{uniform_dist} for various parameters. These PMFs are chosen to represent a wide range of real-world data update scenarios, in the absence of any standard benchmarking dataset (see \cite{dedup}). The truncated exponential PMFs generate thick concentration for lower sparsity levels, yielding best cases for the differential encodings. The uniform distributions illustrate the benefits of the proposed methods for update patterns with no bias on sparse values. The Binomial distributions provide narrow and bell shaped mass functions concentrated around different sparsity levels. The Poisson PMFs model sparse updates spread over the entire support and concentrated around the center. 
 
For a given PMF $\mbox{P}_{\Gamma}(\gamma)$, the average storage size for storing the first two versions is $\mathbb{E}[\delta(\mathbf{x}_{1}, \mathbf{x}_{2})] = n + \sum_{\gamma = 1}^{k} \mbox{P}_{\Gamma}(\gamma) \mbox{min}(2\gamma \kappa, n)$ where $n = \kappa k$. Similarly, the average number of I/O reads to access the first two versions is $\mathbb{E}[\eta(\mathbf{x}_{1}, \mathbf{x}_{2})] = k + \sum_{\gamma = 1}^{k} \mbox{P}_{\Gamma}(\gamma) \mbox{min}(2\gamma, k)$. When compared to the non-differential method, the average percentage reduction in the I/O reads and the average percentage reduction in the storage size are respectively computed as 
\begin{equation}
\label{eq:avg_perc_reads}
\frac{2k - \mathbb{E}[\eta(\mathbf{x}_{1}, \mathbf{x}_{2})]}{2k} \times 100 \mbox{ and } \frac{2n - \mathbb{E}[\delta(\mathbf{x}_{1}, \mathbf{x}_{2})]}{2n} \times 100.
\end{equation}
Since $\delta(\mathbf{x}_{1}, \mathbf{x}_{2}) = \kappa \eta(\mathbf{x}_{1}, \mathbf{x}_{2})$ and $\kappa$ is a constant, the numbers in \eqref{eq:avg_perc_reads} are identical. In Fig. \ref{fig_2}, we plot the percentage reduction in the above quantities for the PMFs displayed in Fig. \ref{fig_1}. The plots show a significant reduction in the I/O reads (and the storage size) when the distributions are skewed towards smaller $\gamma$. However, as expected, the reduction is marginal otherwise. For uniform distribution on $\Gamma$, the plot shows that the advantage with the differential technique saturates for large values of $k$.

\begin{figure}
\begin{center}
\includegraphics[width=3.8in]{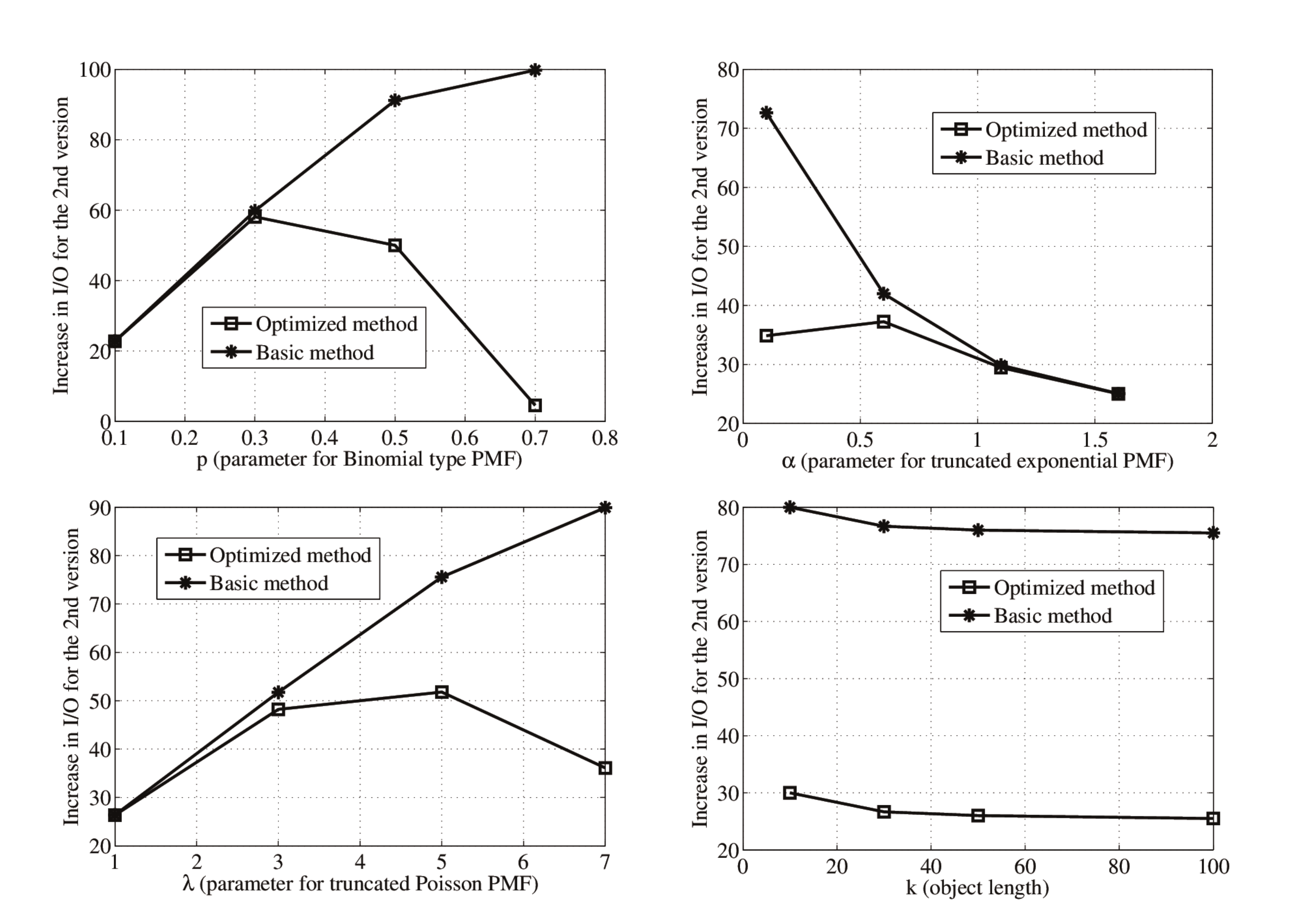}
\vspace{-0.7cm}
\caption{Average percentage increase in the I/O reads to retrieve the 2nd version for the PMFs (in the same order) in Fig. \ref{fig_1} when $L = 2$. The corresponding  values of $n$ and $k$ are same as that of Fig. \ref{fig_2}.}
\label{fig_3}
\end{center}
\end{figure}

We have discussed how the differential technique reduces the storage space at the cost of increased number of I/O reads for the latest version (here the 2nd version) when compared to the non-differential method. For the basic differential encoding, the average number of I/O reads to retrieve the 2nd version is $\mathbb{E}[\eta(\mathbf{x}_{2})] = \mathbb{E}[\eta(\mathbf{x}_{1}, \mathbf{x}_{2})]$. However, for the optimized encoding, $\mathbb{E}[\eta(\mathbf{x}_{2})] = \sum_{\gamma = 1}^{k} \mbox{P}_{\Gamma}(\gamma) f(\gamma)$ where $f(\gamma) = k + 2\gamma$ when $\gamma < \frac{k}{2}$, and $f(\gamma) = k$, otherwise. When compared to the non-differential method, we compute the average percentage increase in the I/O reads for retrieving the 2nd version for both the basic and the optimized methods. Numbers for
\begin{equation}
\label{eq:avg_perc_reads_2nd_ver}
\frac{\mathbb{E}[\eta(\mathbf{x}_{2})] - k}{k} \times 100,
\end{equation}
are shown in Fig. \ref{fig_3}, which shows that the optimized method reduces the excess number of I/O reads for the 2nd version.  
\begin{center}
\begin{table}
\caption{Optimal threshold value for various PMFs with $k = 10$.
\label{table2}
}
\begin{tabular}{|c|c|c|c|c|c|}
\hline
\multicolumn{6}{c}{Binomial: $k=20$, for $\tfrac{k}{2}$-level: $\eta = 40$ and $\delta = 80$}\\
\hline $p$ & $T_{\small{\mbox{opt}}}$ & $\mathbb{E}[\eta]$ & $\mathbb{E}[\delta]$ & $\mathbb{E}[\eta]$ & $\mathbb{E}[\delta]$ \\
&  & (2-level) & (2-level) & ($\frac{k}{2}$-level) & ($\frac{k}{2}$-level) \\
\hline 0.1 & 3  & 28.11 & 56.23 & 24.55 & 49.10 \\
\hline 0.3 & 6  & 35.13 & 70.27 & 31.96 & 63.92 \\
\hline 0.5 & 8  & 38.99 & 77.98 & 38.23 & 76.47 \\
\hline 0.7 & 9  & 39.96 & 79.93 & 39.95 & 79.90 \\
\hline
\multicolumn{6}{c}{Truncated Exponential: $k=10$, for $\tfrac{k}{2}$-level: $\eta = 20$ and $\delta = 40$}\\
\hline $\alpha$ & $T_{\small{\mbox{opt}}}$ & $\mathbb{E}[\eta]$ & $\mathbb{E}[\delta]$ & $\mathbb{E}[\eta]$ & $\mathbb{E}[\delta]$ \\
&  & (2-level) & (2-level) & ($\frac{k}{2}$-level) & ($\frac{k}{2}$-level) \\
\hline 1.6 & 1  &  13.61 &  27.23 & 12.50 & 25.01 \\
\hline 1.1 & 1  &  14.66 &  29.32 & 12.98 & 25.97 \\
\hline 0.6 & 2  &  15.79 &  31.59 & 14.19 & 28.39 \\
\hline 0.1 & 2  &  18.27 &  36.55 & 17.26 & 34.52 \\
\hline
\multicolumn{6}{c}{Truncated Poisson: $k=12$, for $\tfrac{k}{2}$-level: $\eta = 24$ and $\delta = 48$}\\
\hline $\lambda$ & $T_{\small{\mbox{opt}}}$ & $\mathbb{E}[\eta]$ & $\mathbb{E}[\delta]$ & $\mathbb{E}[\eta]$ & $\mathbb{E}[\delta]$ \\
&  & (2-level) & (2-level) & ($\frac{k}{2}$-level) & ($\frac{k}{2}$-level) \\
\hline 1 & 2 & 17.01 & 34.03 & 15.16 & 30.32 \\
\hline 3 & 3 & 20.22 & 40.45 & 18.20 & 36.41 \\
\hline 5 & 4 & 22.24 & 44.49 & 21.06 & 42.13 \\
\hline 7 & 4 & 23.29 & 46.58 & 22.79 & 45.58 \\
\hline
\end{tabular}
\end{table}
\end{center}
\subsection{Two-Level DEC: Threshold Design for $L = 2$ Versions}

We now present simulation results to choose the threshold parameter $1 \leq T \leq \lfloor \frac{k}{2} \rfloor$ for the two-level DEC scheme in Section \ref{sec4_subsec4}. The optimization problem is given in \eqref{th_prob} where
\begin{eqnarray*}
\mathbb{E}[\eta(\mathbf{x}_{1}, \mathbf{x}_{2})] &=& k + \mbox{P}_{\Gamma}(\gamma \leq T) 2T + \mbox{P}_{\Gamma}(\gamma > T) k,
\end{eqnarray*}
$\mathbb{E}[\delta(\mathbf{x}_{1}, \mathbf{x}_{2})] = \kappa \mathbb{E}[\eta(\mathbf{x}_{1}, \mathbf{x}_{2})]$ and $0 \leq w \leq 1$. Since $\mathbb{E}[\delta(\mathbf{x}_{1}, \mathbf{x}_{2})]$ and $\mathbb{E}[\eta(\mathbf{x}_{1}, \mathbf{x}_{2})]$ are proportional, solving \eqref{th_prob} is equivalent to solving instead
\begin{equation}
\label{th_prob_alternate}
T_{\small{\mbox{opt}}} = \arg min_{1 \leq T \leq \lfloor \frac{k}{2} \rfloor} \mathbb{E}[\delta(\mathbf{x}_{1}, \mathbf{x}_{2})].
\end{equation}
In Table \ref{table2}, we list the values of $T_{\small{\mbox{opt}}}$, obtained via exhaustive search over $1 \leq T \leq \lfloor \frac{k}{2} \rfloor$, the average number of I/O reads, the average storage size for the optimized two-level DEC scheme and the $\frac{k}{2}$-level DEC scheme. We denote $\mathbb{E}[\eta(\mathbf{x}_{1}, \mathbf{x}_{2})]$ and $\mathbb{E}[\delta(\mathbf{x}_{1}, \mathbf{x}_{2})]$ by $\mathbb{E}[\eta]$ and $\mathbb{E}[\delta]$, respectively. To compute the average storage size, we use $\kappa = 2$. We see that switching to just two levels of compression incurs negligible loss in the I/O reads (or storage size) when compared to the $\frac{k}{2}$-level DEC scheme. Thus the two-level DEC scheme is a practical solution to reap the benefits of the differential erasure coding strategy.

When Cauchy matrices are used for $\Phi_{T}$, \eqref{th_prob} has to be solved for both \vspace{-2mm}
\begin{eqnarray*}
\mathbb{E}[\eta(\mathbf{x}_{1}, \mathbf{x}_{2})] &=& k + \sum_{\gamma = 1}^{T} \mbox{P}_{\Gamma}(\gamma \leq \gamma) 2\gamma + \mbox{P}_{\Gamma}(\gamma > T) k \\
\mathbb{E}[\delta(\mathbf{x}_{1}, \mathbf{x}_{2})] &=& n + \mbox{P}_{\Gamma}(\gamma \leq T) 2T \kappa + \mbox{P}_{\Gamma}(\gamma > T) k \kappa.
\end{eqnarray*}
Unlike the non-Cauchy case, $\mathbb{E}[\eta(\mathbf{x}_{1}, \mathbf{x}_{2})]$ and $\mathbb{E}[\delta(\mathbf{x}_{1}, \mathbf{x}_{2})]$ are no more proportional and $T_{\small{\mbox{opt}}}$ depends on $w$, $0 \leq w \leq 1$. 

\begin{figure}
\begin{center}
\includegraphics[width=3in]{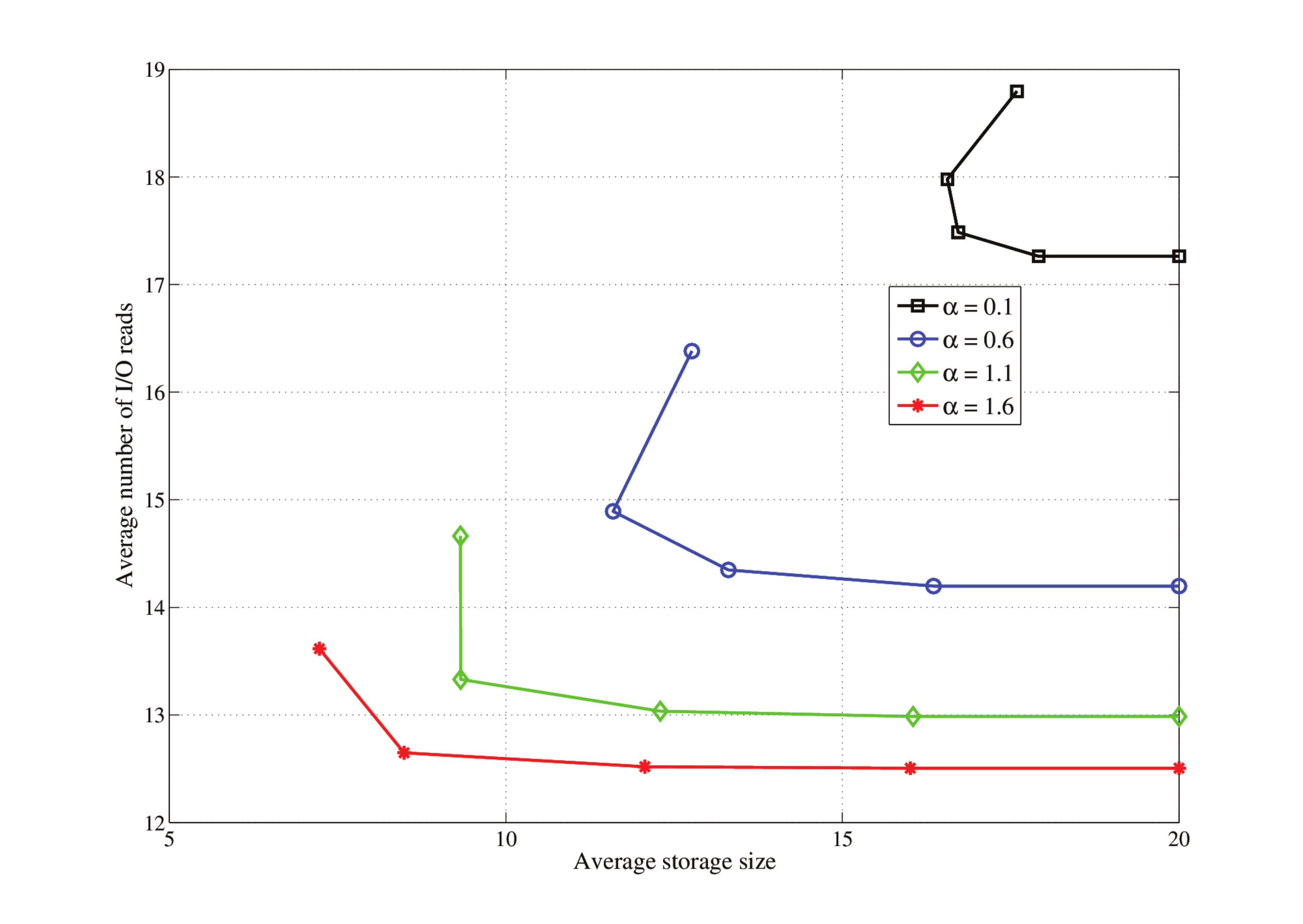}
\vspace{-0.5cm}
\caption{Average storage size $\mathbb{E}[\delta(\mathbf{x}_{1}, \mathbf{x}_{2})]$ versus average number of I/O reads $\mathbb{E}[\eta(\mathbf{x}_{1}, \mathbf{x}_{2})]$, $1 \leq T \leq \lfloor \frac{k}{2} \rfloor = 5$ with truncated exponential distribution.
 For each curve, points from left to right tip correspond to $T = \{1, \ldots, \lfloor \frac{k}{2} \rfloor = 5 \}$.}
\label{fig_8}
\vspace{-3mm}
\end{center}
\end{figure}

To capture the dependency on $w$, we study the relation between $\mathbb{E}[\eta(\mathbf{x}_{1}, \mathbf{x}_{2})]$ and $\mathbb{E}[\delta(\mathbf{x}_{1}, \mathbf{x}_{2})]$ for $1\leq T \leq \lfloor \frac{k}{2} \rfloor$. In Fig. \ref{fig_8}, we plot $\{(\mathbb{E}[\delta(\mathbf{x}_{1}, \mathbf{x}_{2})], \mathbb{E}[\eta(\mathbf{x}_{1}, \mathbf{x}_{2})]), 1\leq T \leq \frac{k}{2}\}$ for the exponential PMFs from Section \ref{subsec:pmf}. For each curve there are $\frac{k}{2} = 5$ points corresponding to $T \in \{1, 2, \ldots, 5\}$ in that sequence from left tip to the right one. The plots indicate the value of $T_{\small{\mbox{opt}}}(w)$ for the two extreme values of $w$, i.e., $w = 0$ and $w = 1$. We further study the curve corresponding to $\alpha = 0.6$. If minimizing $\mathbb{E}[\eta(\mathbf{x}_{1}, \mathbf{x}_{2})]$ is most important with no constraint on $\mathbb{E}[\delta(\mathbf{x}_{1}, \mathbf{x}_{2})]$ (i.e., $w = 1$), then choose $T_{\small{\mbox{opt}}}(1) = \frac{k}{2}$. This option results in $\mathbb{E}[\eta(\mathbf{x}_{1}, \mathbf{x}_{2})]$ which is as low as for the $\frac{k}{2}$-level DEC scheme. While if minimizing $\mathbb{E}[\delta(\mathbf{x}_{1}, \mathbf{x}_{2})]$ is most important with no constraint on $\mathbb{E}[\eta(\mathbf{x}_{1}, \mathbf{x}_{2})]$ (i.e., $w = 0$), then $T_{\small{\mbox{opt}}}(0) = 2$ results in $\mathbb{E}[\delta(\mathbf{x}_{1}, \mathbf{x}_{2})]$ which is the same as for the $2$-level DEC scheme with non-Cauchy matrix. For other values of $w$, the optimal value depends on whether $w > 0.5$. It can be found via exhaustive search over $1 \leq T \leq \lfloor \frac{k}{2} \rfloor$. In summary, using Cauchy matrix for $\Phi_{T}$ reduces the average number of I/O reads to that of the $\frac{k}{2}$-level DEC with just two levels of compression. 

\begin{figure}
\begin{center}
\includegraphics[width=3.8in]{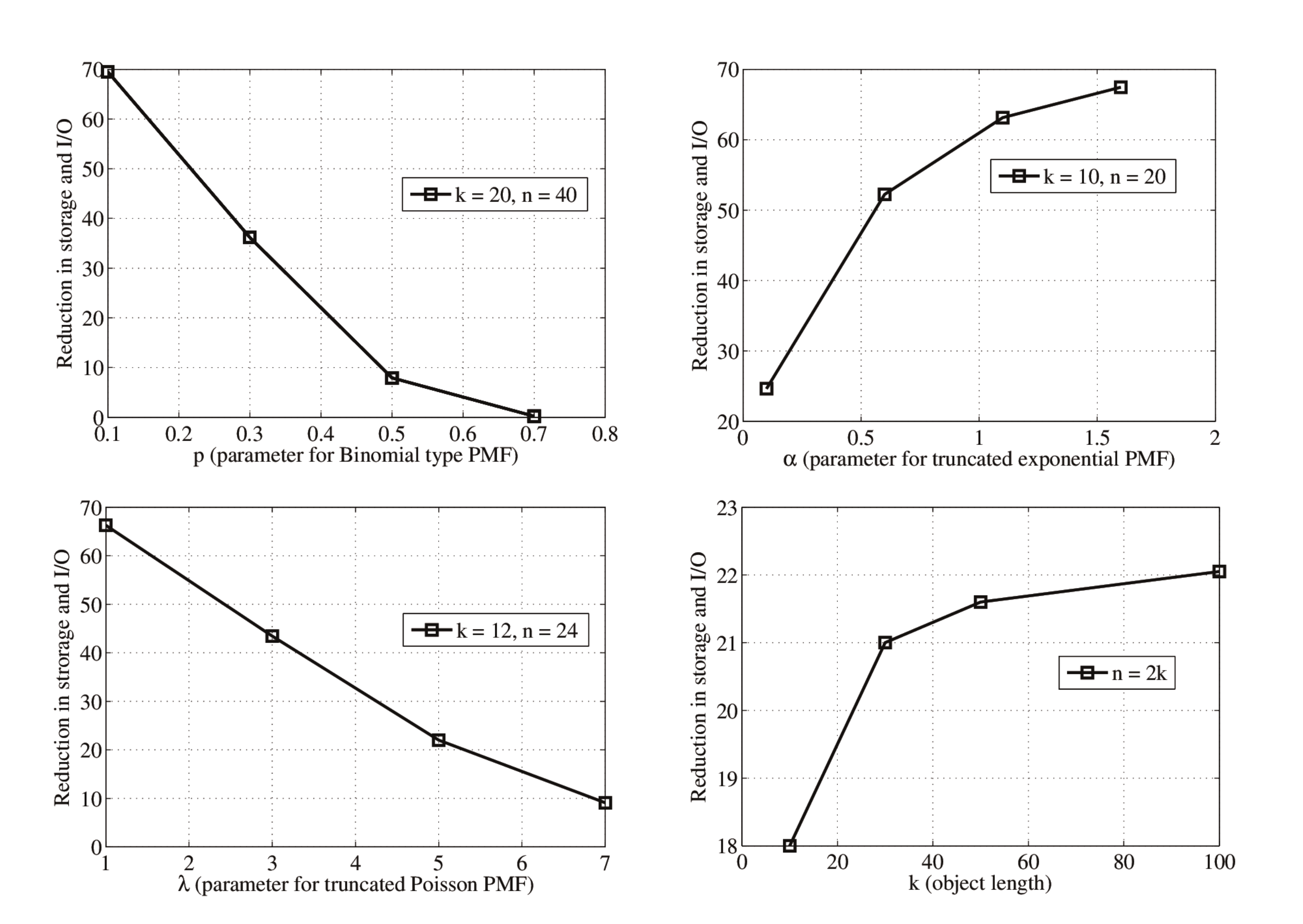}
\vspace{-0.5cm}
\caption{Average percentage reduction in the I/O reads and total storage size for PMFs in Fig. \ref{fig_1} when $L = 10$. The experimental results are presented in the same order as that of the PMFs in Fig. \ref{fig_1}. Identical PMFs are used for the random variable $\{\Gamma_{j}, 2 \leq j \leq 10\}$ to obtain the results.}
\label{fig_L_20}
\vspace{-3mm}
\end{center}
\end{figure}

\subsection{Experimental Results for $L > 2$}
\label{subsec:L>2}

We present the average reduction in the total storage size for a differential system with $L = 10$, assuming identical PMFs on the sparsity levels for every version, i.e., $\mbox{P}_{\Gamma_{j}}(\gamma_{j}) = \mbox{P}_{\Gamma}(\gamma)$ for each $2 \leq j \leq 10$. The average percentage reduction in the total storage size and total I/O reads number are computed similarly to \eqref{eq:avg_perc_reads}, and are illustrated in Fig. \ref{fig_L_20}. The plots show further increase in storage savings compared to $L = 2$ case. In reality, the PMFs across different versions may be different and possibly correlated. These results are thus only indicative of the saving magnitude for storing many versions differentially.

\begin{figure}
\begin{center}
\includegraphics[width=3.8in]{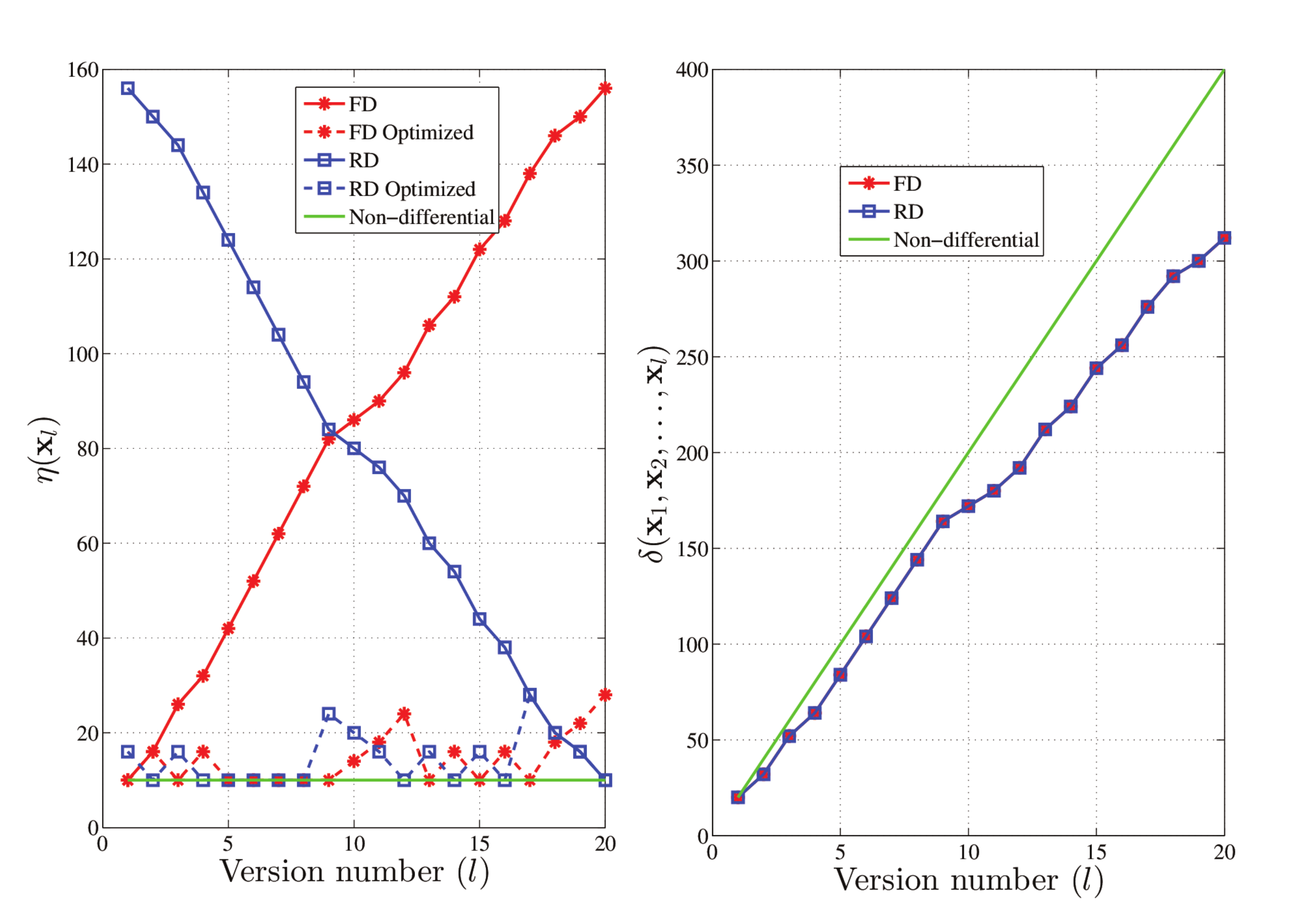}
\vspace{-0.5cm}
\caption{I/O and storage for Examples \ref{ex:L20}-\ref{ex:L20:reverse}. The left plots provide the number of I/O reads to retrieve only the $l$-th version for $1 \leq l \leq 20$. The right plots show the total storage size till the $l$-th version for $1 \leq l \leq 20$. Results are for forward and reverse differential methods, with basic and optimized encoding.}
\label{fig_example_L_20}
\vspace{-3mm}
\end{center}
\end{figure}

\indent To get better insights for $L>2$, in Fig. \ref{fig_example_L_20}, we plot the I/O numbers of Example \ref{ex:L20} and \ref{ex:L20:reverse} for $L = 20$. More than 20\% storage space is saved with respect to the non-differential scheme, for only slightly higher I/O for the optimized DEC.  

%
%

\section{Practical Differential Erasure Coding}
\label{sec:pdec}

So far, we developed a theoretical framework for the DEC scheme under a \emph{fixed object length} assumption across successive versions of the data object (see \eqref{update_eq}). This assumption typically does not hold in practice because of insertions and deletions, which impact the length of the updated object. In this section, we explain how to control zero pads in the file structure so as to support insertions and deletions in a file, while marginally impacting the storage overheads: the variable object size DEC brings a gain of 20\% to 60\% in storage overhead with respect to naive techniques.

\begin{figure}
\includegraphics[width = 3.7in]{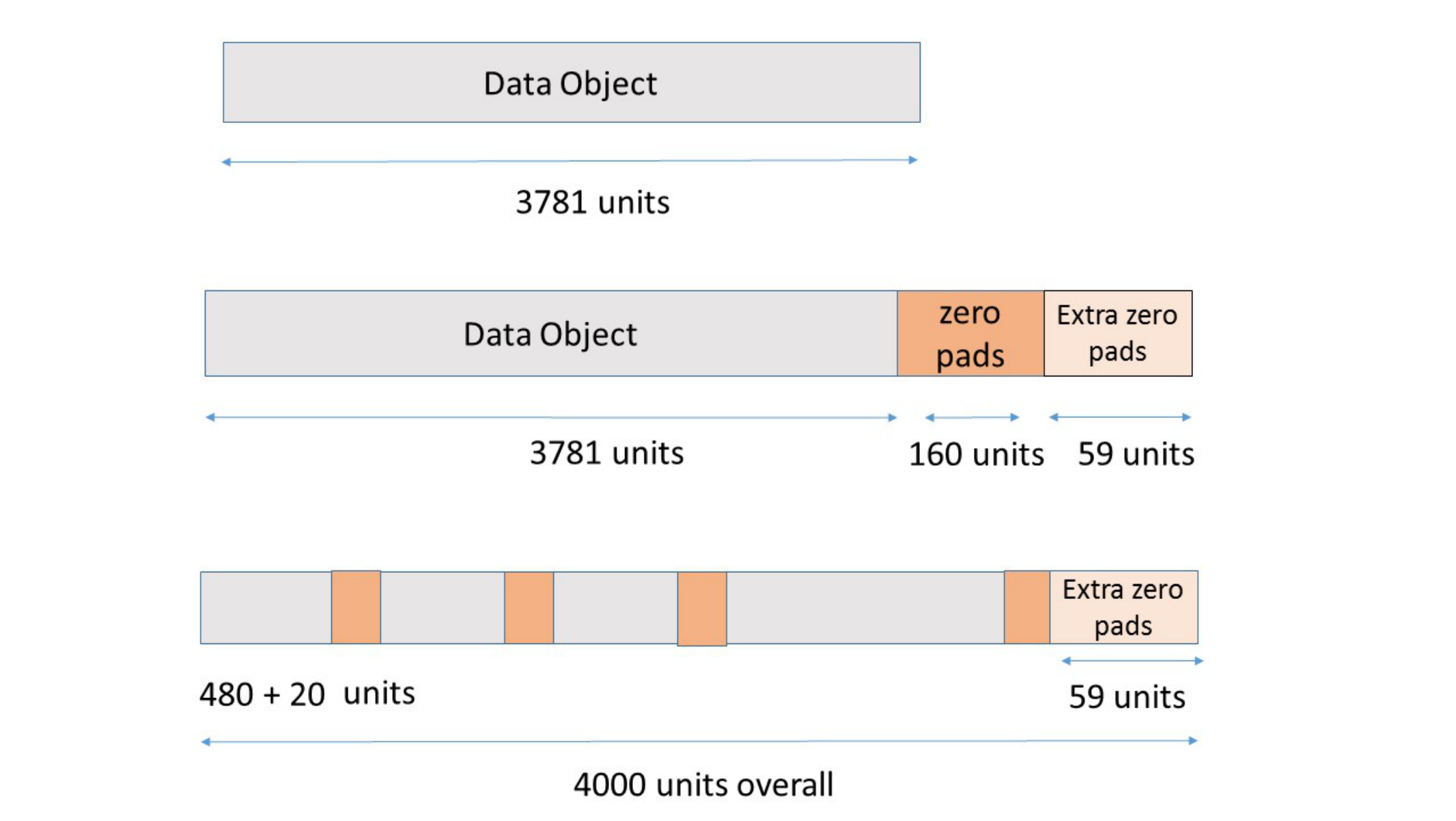}
\vspace{-0.7cm}
\caption{File structure with different placements of zero pads (ZP) - (i) \emph{ZP End} where the zero pads are concentrated at the end (middle figure), and (ii) \emph{ZP Intermediate} where the zero pads are distributed across the file (bottom figure). } 
\label{fig:zero_pads}
\end{figure}

To exemplify the use of zero pads, consider storing a digital object of size 3781 units through a $(12, 8)$ erasure code of symbol size 500 units, as shown in Fig. \ref{fig:zero_pads}. Since the object is encoded blockwise, $219$ zero pads are added to extend the object size to $4000$ units. The zero pads naturally absorb insertions made anywhere in the file, as long as the total size is less than $219$ units, thus retaining the length of the updated version to $4000$ units. However, since the zero pads are placed at the end, insertions made at the beginning of the file propagate changes across the rest of the file. The difference object is thus unlikely to exhibit sparsity.
Alternatively, one could distribute zero pads across the file at different places as shown in Fig. \ref{fig:zero_pads}. Here 160 zero pads are distributed at 8 patches with each patch containing 20 zero pads. This strategy arrests propagation of changes when (small size) insertions are made either at the beginning or middle of the file. 

Despite zero padding looking like a natural way to handle insertions, it is already clear from this example that the optimization of the size and placements of zero pads is not immediate. We defer this analysis to Section \ref{sec7}, and firstly emphasize  
the functioning of the variable size DEC scheme.

\subsection{DEC \textbf{Step 1} for Variable Size Length Object}

Let $\mathcal{F}_{1}$ be the first version of a file of size $V$ units. The system distributes the file contents into several \emph{chunks}, each of size $\Delta$ units. Within each chunk, $\delta < \Delta$ units of zero pads are allocated at the end while the rest of it are dedicated for the file content. 
Thus, the $V$ units of the file are spread across 
\begin{equation}
\label{eq:nr_chunks}
M = \left\lceil \frac{V}{\Delta - \delta}\right\rceil
\end{equation}
chunks $\{C_{1}, C_{2}, \ldots, C_{M}\}$, where $\lceil \cdot \rceil$ denotes the ceiling operator. The zero pads added at the end of every chunk promote sparsity in the difference between two successive versions. 

Once the file contents are divided into $M$ chunks, they are stored across different servers, using an $(n,k)$ erasure code: the code is applied on a block of $k$ data chunks to output $n (> k)$ chunks which includes the data chunks and $n-k$ encoded chunks that are generated to provide fault tolerance against potential failures. The parameter $k$ is optimized for the architecture with respect to $M$, which is file dependent:

{\bf Case 1:} When $M < k$, additional $M-k$ chunks containing zeros are appended to create a block of $k$ chunks. Henceforth, these additional chunks are referred to as \emph{zero chunks}. Then, the $k$ chunks are encoded using an $(n, k)$ erasure code. 

{\bf Case 2:} When $M \geq k$, the $M$ chunks are divided into $G = \lceil \frac{M}{k}\rceil$ groups $\mathcal{G}_{1}, \mathcal{G}_{2}, \ldots, \mathcal{G}_{G}$. The last group $\mathcal{G}_{G}$ if found short of $k$ chunks is appended with \emph{zero-chunks}. The $k$ chunks in each group are encoded using an $(n, k)$ erasure code. 

For the first version $\mathcal{F}_1$, the $G$ groups of chunks together have $\delta M + N\Delta$ units of zero pads, where $1\leq N < k$, represents the number of zero-chunks added to make $\mathcal{G}_{G}$ contain $k$ chunks. In addition, the $M$-th chunk may have extra padding due to the rounding operation in \eqref{eq:nr_chunks}. The $\delta M$ units of zero pads that are distributed across the chunks shield propagation of changes across chunks when an insertion is made in subsequent file versions. This object can now withstand a total of $\delta M$ units of insertion (anywhere in the file if $\delta M < N\Delta$) by retaining $G$ groups for the second version. 



We next discuss the use of zero pads while storing the $(j+1)$-th version $\mathcal{F}_{j+1}$ of the file, $j \geq 1$.

\subsection{DEC \textbf{Step} $j+1$ under Insertions}
\label{sec3_subsec1}

For the $(j+1)$-th version, the DEC system is designed to identify the difference in the file content size in every chunk. Then the changes in the file contents are carefully updated in the chunks, in the increasing order of the indices $1, 2, \ldots, M$, so as to minimize the number of chunks modified due to changes in one chunk. For $1 \leq i \leq M$, if the content of $C_{i}$ grows in size by at most $\delta$ units, then some zero pads are removed to make space for the expansion. This $C_{i}$ will have fewer zero pads than the first version. On the other hand, if the content of $C_{i}$ grows in size by more than $\delta$ units, then the first $\Delta$ units of the file content are written to $C_{i}$ while the remaining units are shifted to $C_{i+1}$. The existing content of $C_{i+1}$ is in turn shifted, and hence, it will have fewer zero pads than $\delta$. The propagation of changes in the chunks continue until all the changes in the file are reflected.

\subsection{DEC \textbf{Step} $j+1$ under Deletions}
\label{sec3_subsec2}

We saw that placing zero pads stops propagation of changes across chunks when inserting new contents. When file contents are deleted, the zero pads continue to block propagation, this time in the reverse direction. Since deletion results in reduced size of the file contents in chunks, this is equivalent to having additional zero pads (of the same size as that of the deleted patch) in the chunks along with the existing zero pads. After this process, the metadata should reflect the total size of the file contents (potentially less than $\Delta - \delta$) in the modified chunk. Thus, deletion of file contents boosts the capacity of the data structure to shield larger insertions in the next versions. 


\subsection{Encoding Difference Objects}
\label{sec3_subsec3}

The preceding sections emphasized the need for shifting the file contents across the chunks when the insertion size is more than $\delta$ units. Digging further, if the insertion size is large enough, then new chunks (or even new groups) have to be added to the existing chunks (or groups), thus changing the object size of the $(j+1)$-th version. Note that the differential encoding strategy requires two successive versions to have the same object size to compute the difference. In particular, we adopt the reverse DEC wherein the latest version of the object is stored in full while the preceding versions are stored in a differential manner. Once the contents of the $(j+1)$-th version is updated to the chunks, we compute the difference between the chunks of the $j$-th and the $(j+1)$-th version. Then we declare a difference chunk to be non-zero if it contains at least one non-zero element. Within a group, if the number of non-zero chunks, say $\gamma$ of them, is smaller than $\frac{k}{2}$ then the difference object is compressed to contain $2\gamma$ chunks, before encoding them using a $\frac{k}{2}$-level DEC scheme discussed in Section \ref{sec2}. We continue this procedure of storing the difference objects until the modified object size is at most $kG$ chunks. Note that one could also use the two-level DEC scheme in Section \ref{sec4} as an alternate option to store the difference objects with just two erasure codes.

A set of consecutive versions of the file that maintains the same number of groups is referred to as a \emph{batch} of versions, while the number of such versions within the batch is called the depth of the batch. The case when insertions change the group size is addressed next as a source for resetting the differential encoding strategy.  



\subsection{Criteria to Reset SEC}
\label{sec3_subsec4}

\textbf{Criterion 1:} Starting from the second version, the process of storing the difference objects continues until $G$ remains constant. When the changes require more than $G$ groups, i.e., the updates require more than $kG$ chunks, the system terminates the current batch, and then stores the object in full by redistributing the file contents into a new set of chunks. To illustrate this, let the $j$-th version of the file (for some $j > 1$) be distributed across $M_{j}$ chunks, where $\lceil \frac{M_{j}}{k} \rceil \leq G$. Now, let the changes made to the $(j+1)$-th version occupy $M_{j+1}$ chunks where $\lceil \frac{M_{j}}{k} \rceil > G$. At this juncture, we reorganize the file contents across several chunks with $\delta$ units for zero pads (as done for the first version). After re-initialization, this file has $G' = \lceil \frac{M_{j+1}}{k} \rceil$ groups. 

\textbf{Criterion 2:} Another criterion to reset is when the number of non-zero chunks is at least $\frac{k}{2}$ within every group. Due to insufficient sparsity in each group, there would be no saving in storage size in this case, and as a result, a new batch has to be started. However, a key difference from criterion 1 is that the contents of the chunks are not reorganized since the group size has not changed.

\section{Experiments with Practical DEC}
\label{sec7}


We conduct experiments with several synthetic workloads, capturing wide spectrum of realistic loads to demonstrate the efficacy of our scheme. The main objectives are 
\begin{enumerate}
\item to determine the right strategy to place the zero pads in order to promote sufficient sparsity in the difference object for different classes of workloads (see Sections \ref{sec7_subsec1} and \ref{sec7_subsec2}).
\item to compare the storage savings of DEC against two baselines, namely (i) \AD{a system setup using} concepts from Rsync, which is fundamentally a delta encoding technique for file transfer and file synchronization systems, and (ii) a naive technique where each version is fully coded and treated as distinct objects, referred to as non-differential scheme (see Section \ref{sec7_subsec3}).
\end{enumerate}

Throughout this section, we use the reverse differential method where the order of storing the difference vectors is reversed as $\{\mathbf{z}_{2}, \mathbf{z}_{3}, \ldots, \mathbf{z}_{L}, \mathbf{x}_{L}\}$, as it facilitates direct access to the latest version of the object. Also, DEC scheme refers not to the primitive form discussed in Section \ref{sec2}, \AD{but instead it} refers to its variant which was discussed in Section \ref{sec:pdec}. Unless specified \AD{otherwise}, we showcase only the best case storage benefits that come with the application of $\frac{k}{2}$-level DEC scheme, wherein the $\frac{k}{2}$ erasure codes are assumed to have identical storage overhead of $\kappa = 2$.

For the DEC scheme storing two versions, i.e., $L = 2$, the average storage size for the second version is given by 
\begin{equation}
\label{eq:storage_size_ec}
\mathbb{E}[\delta(\mathbf{z}_{2})] = \kappa \mathbb{E}[\mbox{min}(2\gamma_{j}, k)],
\end{equation}
which is the average size of the data object after erasure coding. Since the storage overhead $\kappa$ is held constant for all the $\frac{k}{2}$ erasure codes, we note that the quantity
\begin{equation}
\label{eq:storage_size}
\frac{\mathbb{E}[\delta(\mathbf{z}_{2})]}{\kappa} = \mathbb{E}[\mbox{min}(2\gamma_{j}, k)],
\end{equation}
which is the average storage size prior to erasure coding, is a sufficient statistic to evaluate the placement of zero pads. Henceforth, we use \eqref{eq:storage_size} as the yardstick in our analysis. However, in general, when storage overheads are different, $\mathbb{E}[\delta(\mathbf{z}_{2})]$ in \eqref{eq:storage_size_ec} is a relevant metric for the analysis.

Notice that unlike the quantities in Section \ref{sec:simulation}, the quantity in \eqref{eq:storage_size} includes raw data as well as zero pads. This difference is attributed to a more realistic model of erasure coded versioning system in Section \ref{sec:pdec}, where the zero pads facilitate block encoding of arbitrary sized data objects in addition to shielding the rippling effect from insertions and deletions.

\subsection{Comparing Different Placements of Zero Pads}
\label{sec7_subsec1}

\begin{figure}
\includegraphics[width = 3.6in]{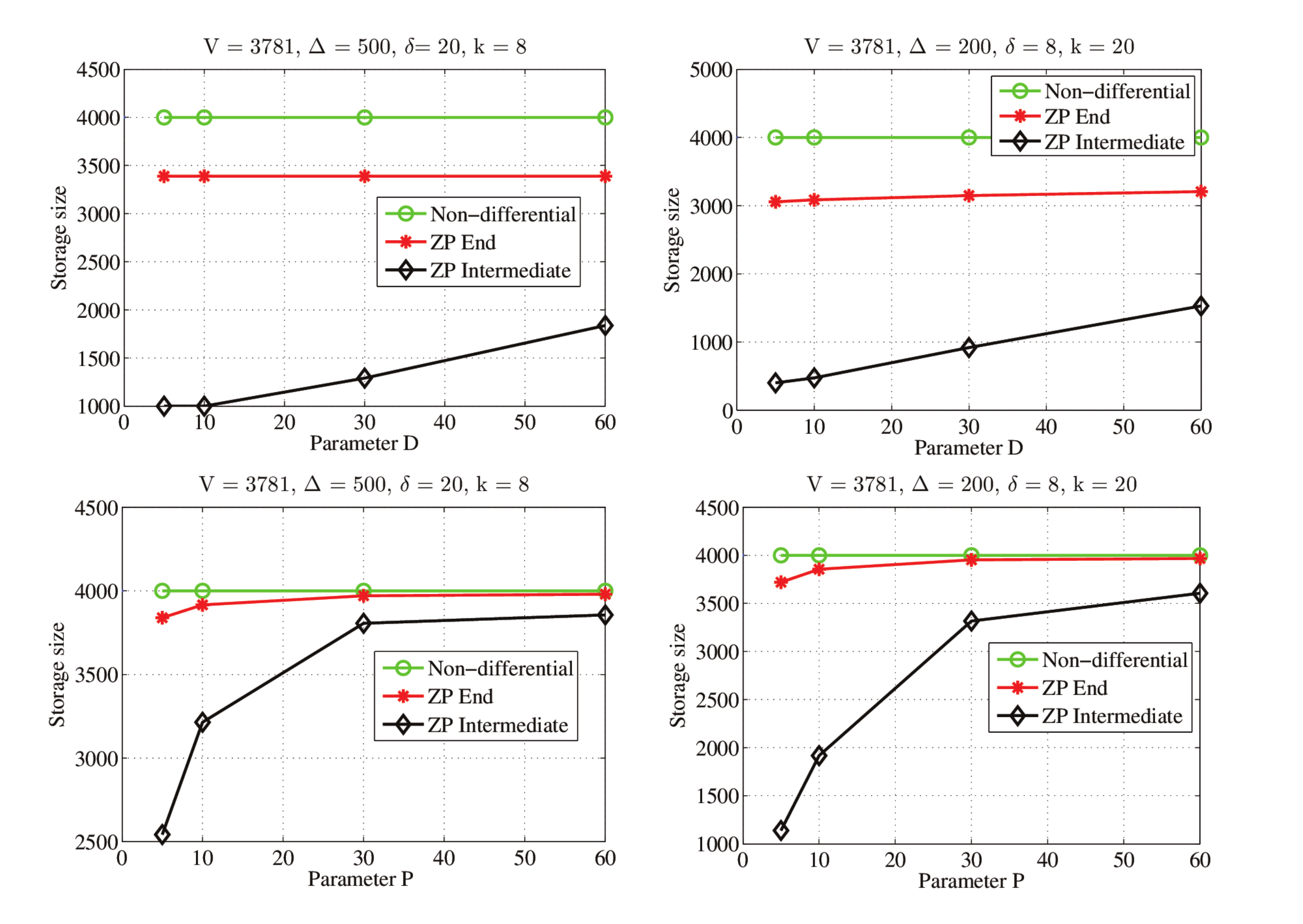}
\vspace{-0.5cm}
\caption{Comparing different placements of zero pads against insertions: Average storage size (as given in \eqref{eq:storage_size}) for the 2nd version against workloads comprising random insertions. For the top figures, workloads are bursty insertions whose size is uniformly distributed in the interval $[1, D]$ for $D \in \{5, 10, 30, 60\}$. For the bottom figures, workloads are several single unit insertions whose quantity is distributed uniformly in the interval $[1, P]$, where $P \in \{5, 10, 30, 60\}$.
\label{fig:zp_insertions}
}
\end{figure}

\begin{figure}
\includegraphics[width = 3.6in]{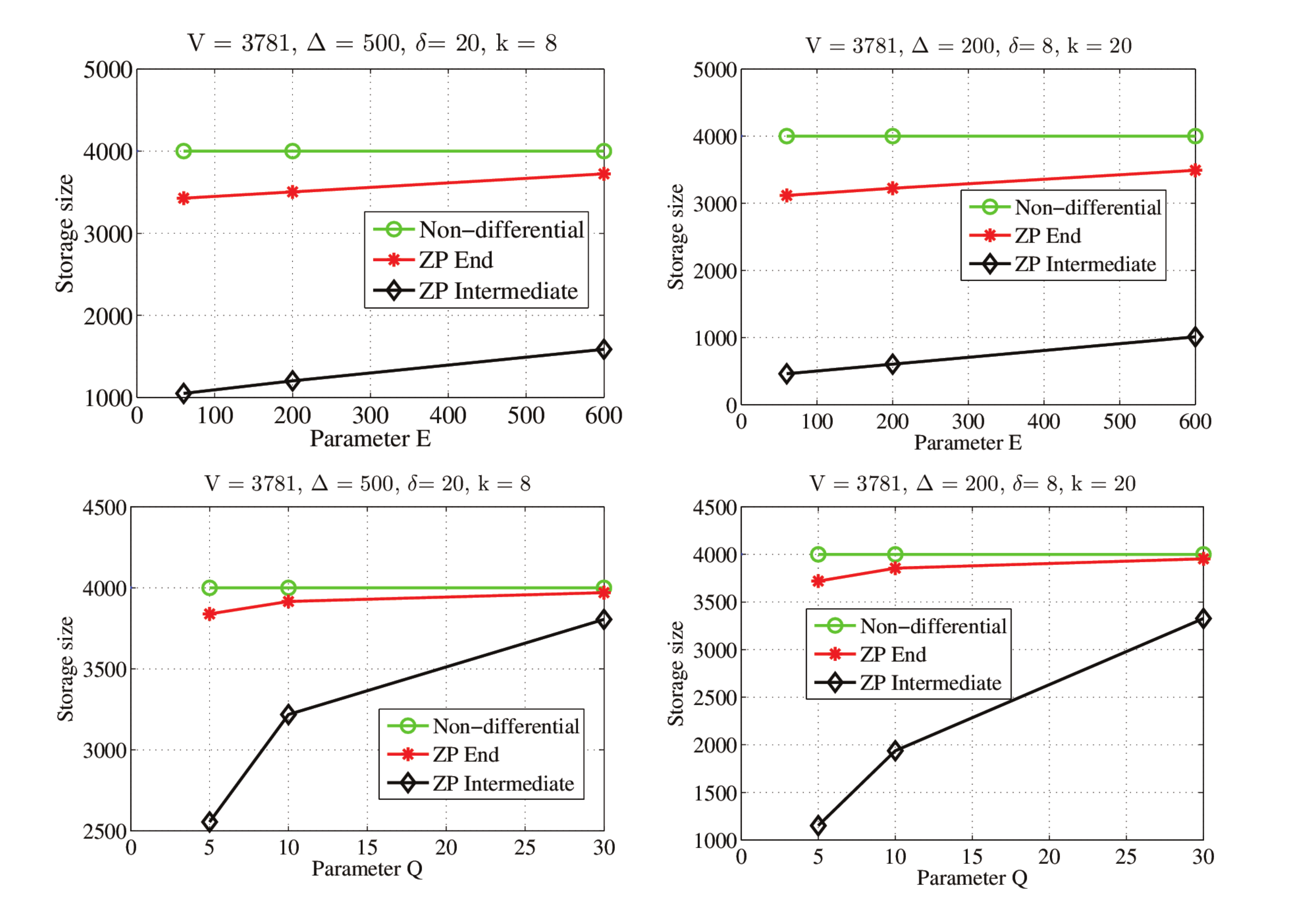}
\vspace{-0.5cm}
\caption{Comparing different placements of zero pads against deletions: Average storage size (as given in \eqref{eq:storage_size}) for the second version against workloads comprising random deletions. For the top figures, workloads are single bursty deletions whose size is uniformly distributed in the interval $[1, E]$ for $E \in \{60, 200, 600\}$.  For the bottom figures, workloads are several single unit deletions whose quantity is distributed uniformly in the interval $[1, Q]$, where $Q \in \{5, 10, 30\}$.
\label{fig:zp_deletions}
}
\end{figure}

We conduct several experiments to compare the storage savings from the zero pads placements highlighted in Fig. \ref{fig:zero_pads}. The parameters for the experiment are $V = 3781$, $\Delta = 500, \delta = 20$ and $k = 8$. The two schemes under comparison are \emph{ZP End} and \emph{ZP Intermediate} (discussed in Fig. \ref{fig:zero_pads}), where the zero pads are allocated at the end and at intermediate positions, respectively. Like \emph{ZP Intermediate} scheme, the \emph{ZP End} scheme also contains $k = 8$ chunks (each of size $\Delta$), however in this case, $219$ zero pads appear at the end in the $8$-th chunk. In general, appending zero pads at the end of the data object is a necessity to employ erasure codes of fixed block length. Thus, for the parameters of our experiment, both the \emph{ZP End} and \emph{ZP Intermediate} schemes initially have equal number of zero pads (but at different positions), and hence, the comparison is fair. 

From our experiments, we compute the average numbers in \eqref{eq:storage_size} when two classes of random insertions are made to the first version, namely: (i) single bursty insertion whose size is uniformly distributed in the interval $[1, D]$, for $D = 5, 10, 30, 60$, and (ii) several single unit insertions uniformly distributed across the object, where the number of insertions is uniformly distributed in the interval $[1, P]$, where $P = 5, 10, 30, 60$. We repeat the experiments 1000 times by generating random insertions and then compute the average storage size of the compressed object $\mathbf{z}'_{2}$ (as given in \eqref{eq:storage_size}).  In Fig. \ref{fig:zp_insertions} we plot the average storage size with the \emph{ZP End} and \emph{ZP Intermediate} schemes. Similar plots are also presented in Fig. \ref{fig:zp_insertions} (on the right) with parameters $\Delta = 200, \delta = 20$ and $k = 20$ for the same object. The plots highlight the advantage of distributing the zero pads as it can arrest the propagation of changes through intermediate zero pads. We conduct more experiments for several classes of random deletions and the results are presented in Fig. \ref{fig:zp_deletions}, which highlight the savings in storage size for the \emph{ZP Intermediate} scheme.

\begin{figure}
\includegraphics[width = 3.7in]{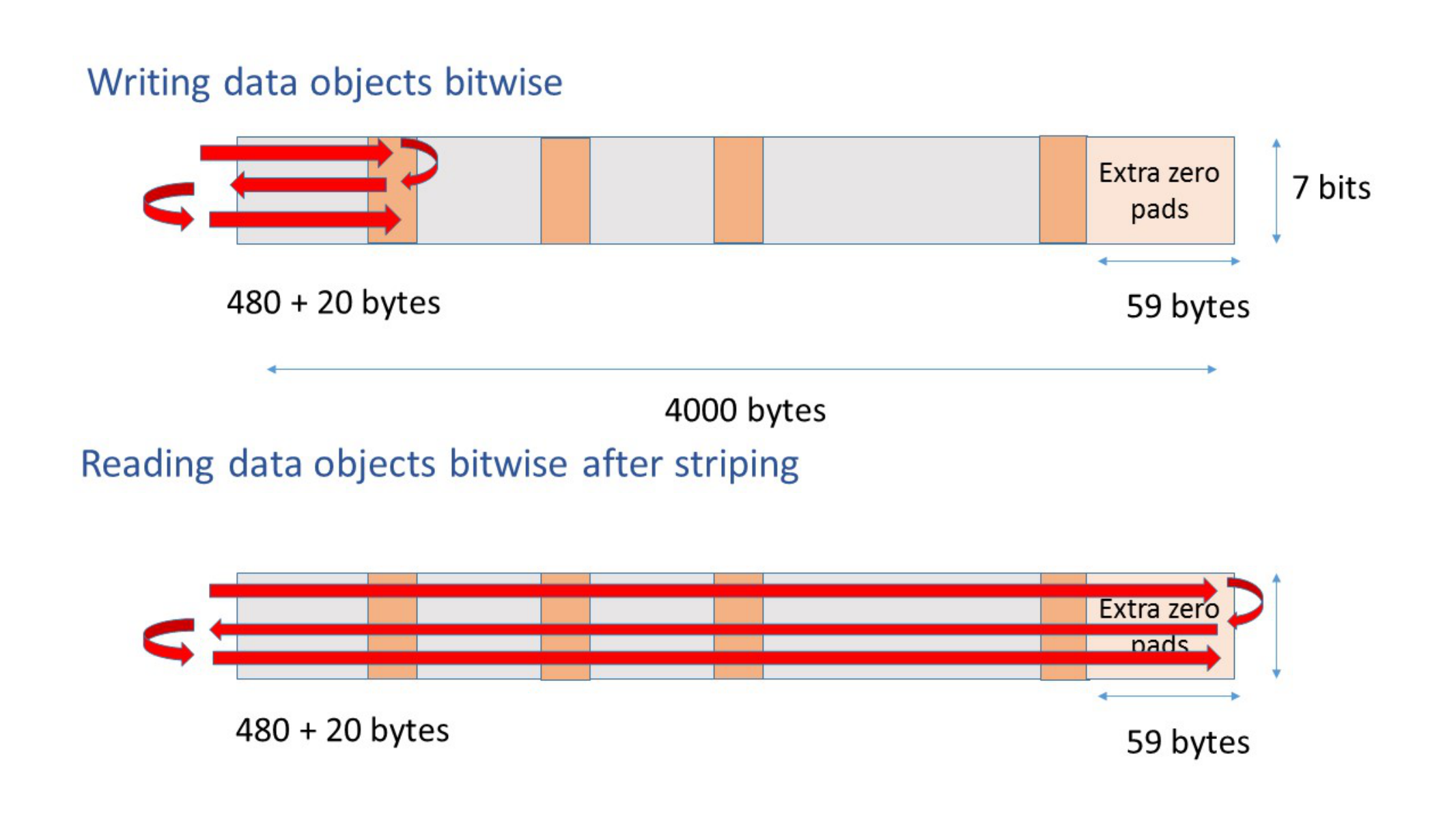}
\vspace{-0.7cm}
\caption{Bit striping method to generate striped chunks. Top figure depicts bit-level writing of data into the chunks. Bottom figure depicts bit-level reading of data. This technique is suitable for uniformly distributed sparse insertions. } 
\label{fig:zero_pads_bit_striping}
\end{figure} 

\begin{figure}
\begin{center}
\includegraphics[scale=0.40]{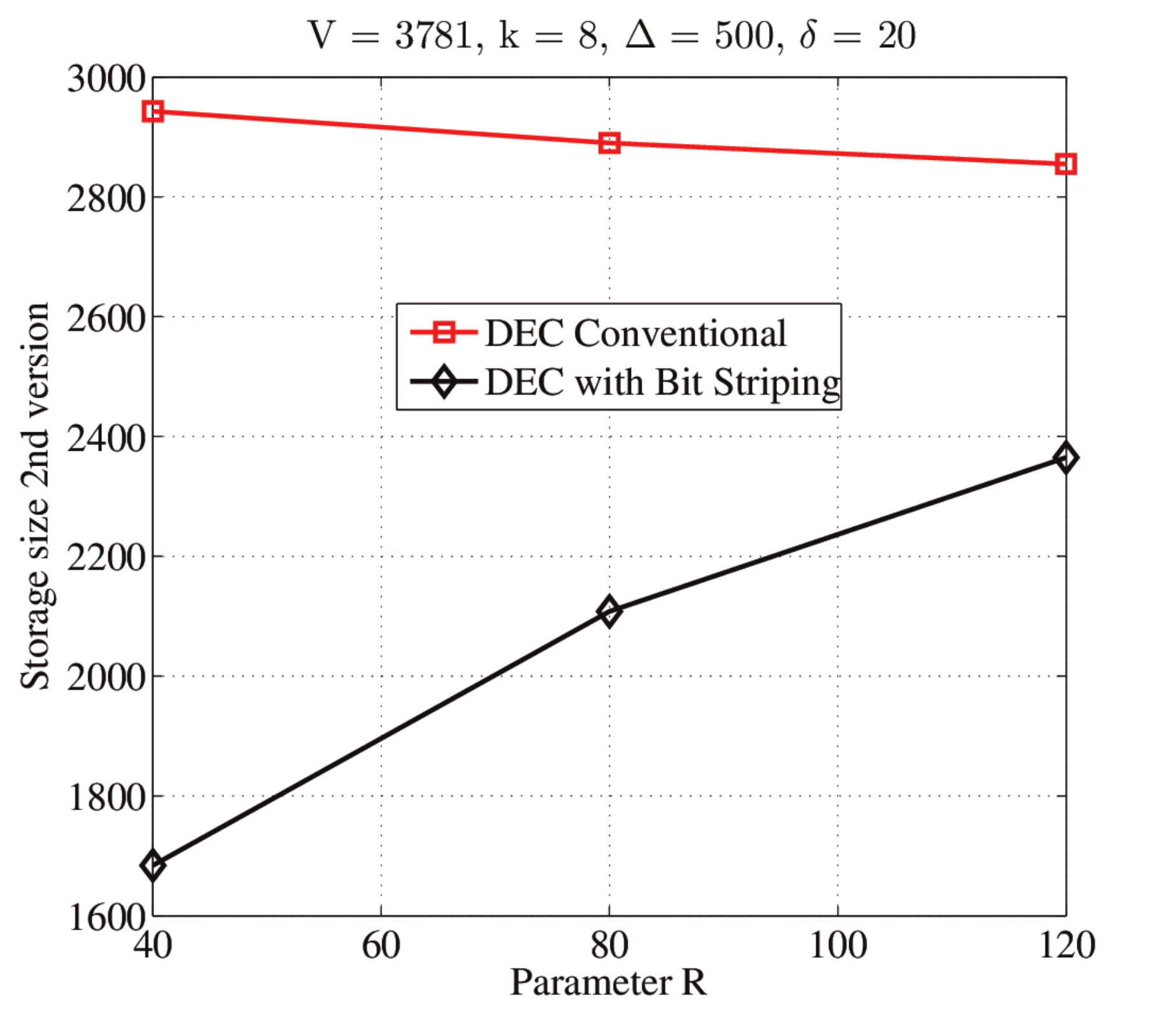}
\end{center}
\vspace{-0.5cm}
\caption{Comparison of DEC schemes with and without bit striping. Average storage size (given in \eqref{eq:storage_size}) for the second version against workload that has 3 single unit insertions with intra-distance uniformly distributed in the interval $[\Delta - \delta - R, \Delta - \delta + R]$, where $R \in \{40, 80, 120\}$. For the experiments, we use $\Delta = 500$ and $\delta = 20$.
\label{fig:sec_bit_striping}
}
\end{figure}

\subsection{Chunks with Bit Striping Strategy}
\label{sec7_subsec2}

In this section, we analyze the right strategy to synthesize chunks for workloads that involve several single insertions with sufficient spacing. We first explain the motivation for this special case using the following toy example. Consider storing a data object of size $V = 3871$ units using the parameters $\Delta = 500, \delta = 20, k = 8$. Assume that $3$ units of insertions are made to the object at the positions $1, 481$ and $961$, which translates to modifications of the chunks $C_{1}, C_{2}$ and $C_{3}$, respectively. Thus, due to just $3$ single unit insertions, three chunks are modified because of which the difference object after compression will be of size $3000$ units. Instead, imagine striping every chunk into $k$ partitions at the bit level such that the $\delta$ zero pads are equally distributed across the partitions (see the top figure in Fig. \ref{fig:zero_pads_bit_striping}). Then, create a new set of $k$ chunks as follows: create the $t$-th chunk for $1 \leq t \leq k$ by concatenating the contents in the $t$-th partition of all the original chunks (see the bottom figure in Fig. \ref{fig:zero_pads_bit_striping}). By applying this striping method to the toy example, we see that only one chunk (after striping) is modified, hence, this strategy would need only $1000$ units for storage after compression.

For the above example, the insertions are spaced exactly at intra-distance $\Delta - \delta$ units to highlight the benefits, although in practice, the insertions can as well be approximately around that distance to reap the benefits. We conduct experiments by introducing 3 random insertions into the file, where the first position is chosen at random while the second and the third are chosen with intra-distance (with respect to the previous insertion) that is uniformly distributed in the interval $[\Delta - \delta - R, \Delta - \delta + R]$ when $R \in \{40, 80, 120\}$. For this experiment, the average storage size for the second version (i.e., the size of the compressed object $\mathbf{z}'_{2}$ given in \eqref{eq:storage_size}) is presented in Fig. \ref{fig:sec_bit_striping}, which shows significant reduction in storage for the striping method when compared to the conventional method. Notice that as $R$ increases, there is higher chance for the neighboring insertions to not fall in the same partition number of different chunks, thus diminishing the gains. 

\begin{figure}
\includegraphics[width = 3.6in]{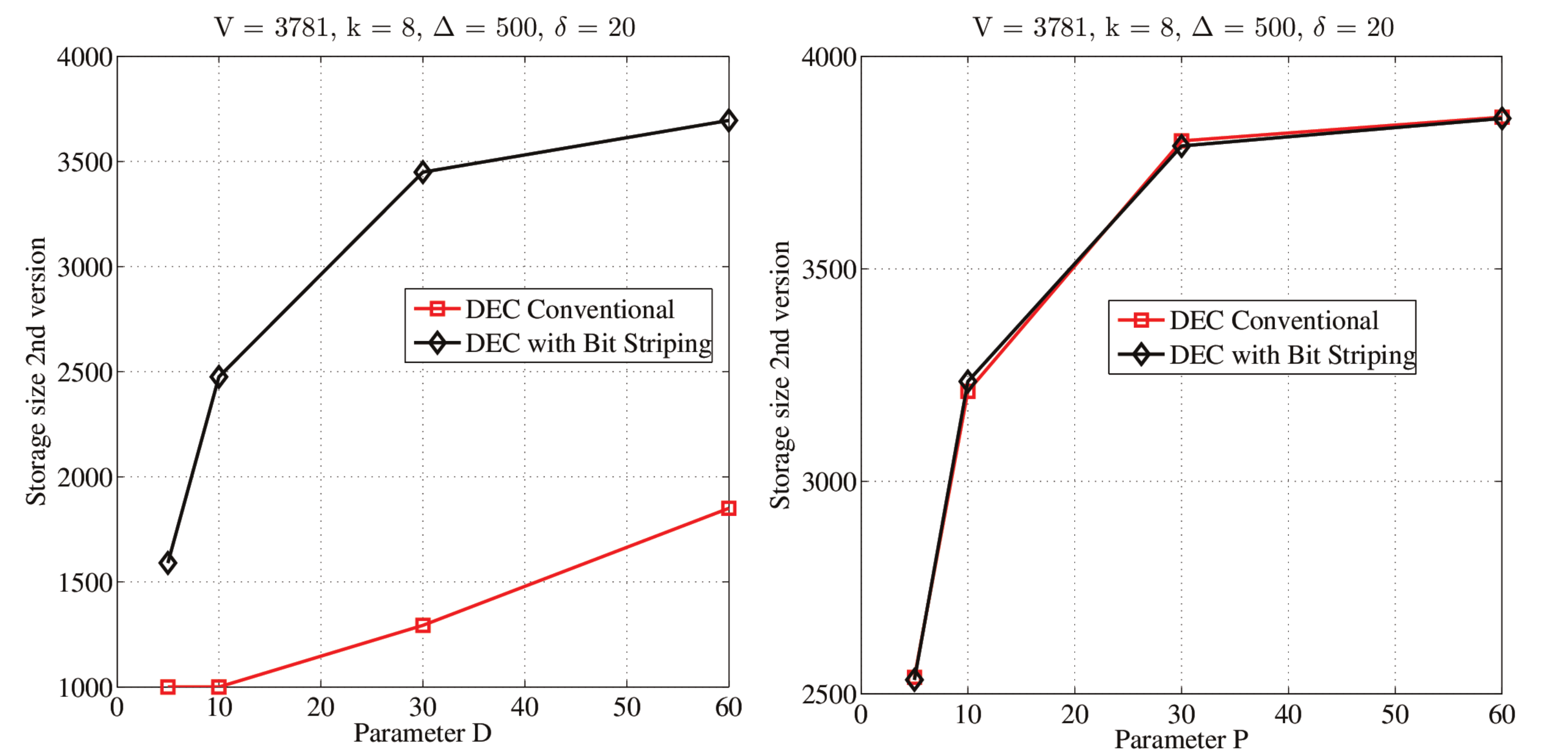}
\vspace{-0.5cm}
\caption{Comparing DEC schemes with and without bit striping against bursty (the left plot) and randomly distributed single insertions (the right plot) with parameters $D, P \in \{5, 10, 30, 60 \}$.
\label{fig:sec_bit_striping_gen}
}
\end{figure}

\indent We also test the striping method against two types of workloads, namely, the bursty insertion (with parameter $D \in \{5, 10, 30, 60\}$) and the randomly distributed single insertions with parameter $P \in \{5, 10, 30, 60\}$. For the workloads with single insertions, the spacing between the insertions is uniformly distributed and not necessarily at intra-distance $\Delta - \delta$. In Fig. \ref{fig:sec_bit_striping_gen}, we present the average storage size for the second version (given in \eqref{eq:storage_size}) against such workloads. The plots show significant loss for the striping method against the former workload (as they are not designed for such patterns), whereas the storage savings are approximately close to the conventional method against the latter workload. In summary, if the insertion pattern is known to be distributed a priori, then we advocate the use of the striping method as it provides similar performance as that of the conventional method with a potential to provide reduced storage savings for some special distributed insertions.

\subsection{Comparing Storage Savings from DEC against Rsync based Distributed Storage System}
\label{sec7_subsec3}


\begin{figure}
\includegraphics[width = 3.6in]{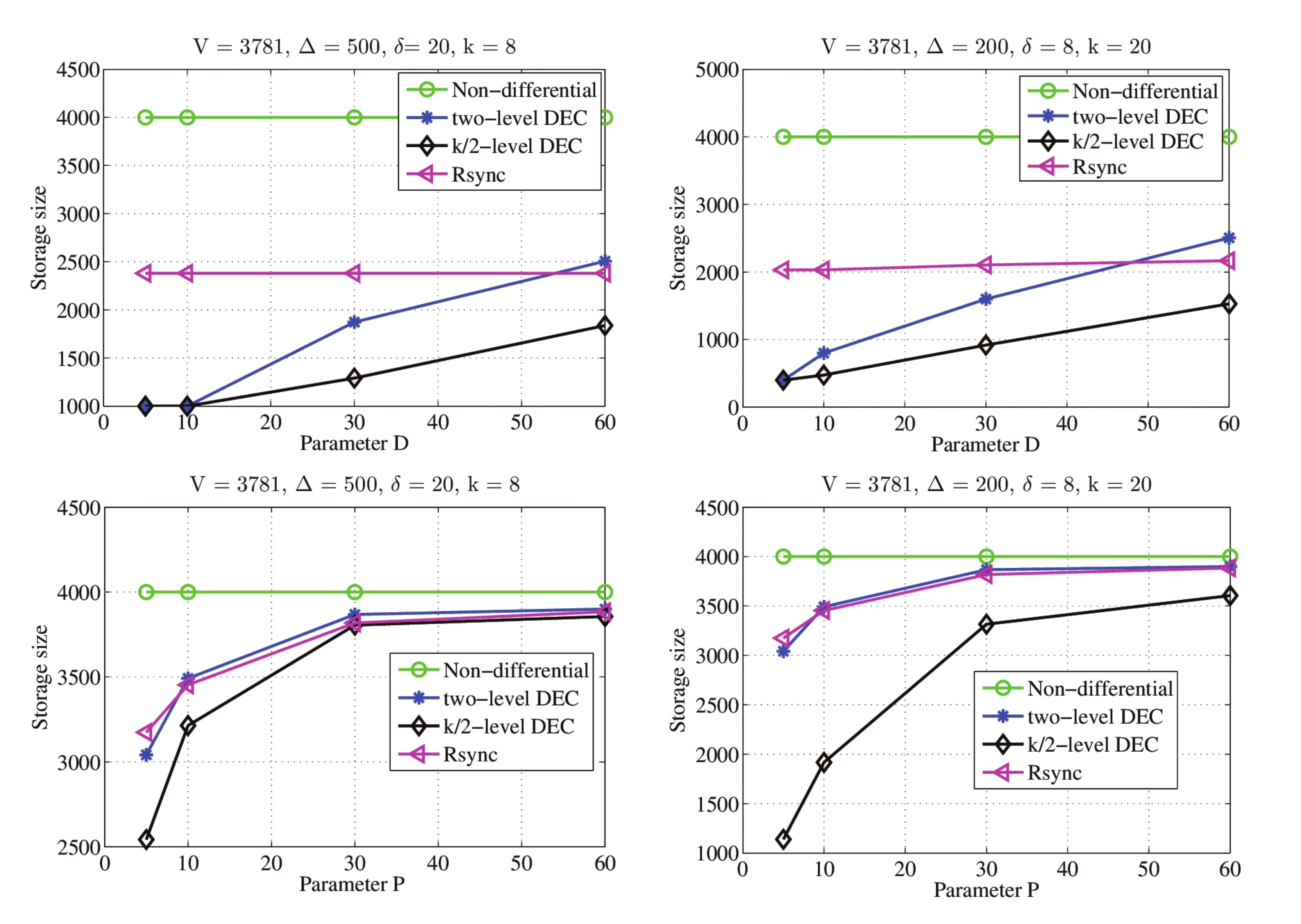}
\vspace{-0.5cm}
\caption{DEC vs. Rsync with respect to insertions: Average storage size for the 2nd version against workloads comprising random insertions. The parameters $D$ and $P$ are as defined for Fig. \ref{fig:zp_insertions}. The $\frac{k}{2}$-level DEC scheme applies an erasure code for each sparsity level, whereas the two-level DEC applies only two erasure codes based on the threshold $T_{\small{\rm{opt}}}$. The left and the right plots are for bursty and distributed single insertions, respectively. 
\label{fig:rsync_vs_dec_insertions}
}
\end{figure}


An important yardstick for comparison is a system \AD{setup using} concepts from Rsync \cite{rsync}, which is a delta encoding technique for file transfer and file synchronization systems. In the original Rsync algorithm, only the modified chunks between the successive versions are transferred across the servers, thereby reducing the communication bandwidth. However, with the application of Rsync ideas to \AD{versioned} storage systems, the gains in the communication bandwidth gets translated to gains in the storage size. In particular, if $\gamma < \frac{k}{2}$ chunks are modified, then only those $\gamma$ modified chunks are stored in the Rsync based scheme by keeping track \AD{of} the indices of the modified chunks, whereas, in contrast, $2\gamma$ chunks are stored in the DEC scheme without storing the indices of the modified chunks and yet able to recover them accurately from the compressed $2\gamma$ chunks. Note that the Rsync based method is effective if the updated version retains its size, and has few in-place modifications. However, if the object size changes due to insertions or deletions, or when changes propagate at bit level, then dividing the updated version into fixed size chunks need not result in sufficient sparsity in the difference object across versions. For the Rsync based method, although there are no preallocated zero pads, they indirectly appear at the end to generate $k$ (or its multiple) number of chunks. 

We conduct more experiments to compare the storage savings offered by DEC and Rsync. This time the parameters of the experiment are $V = 3871$ $\Delta = 500, \delta = 20, k = 8$, and the workload includes random insertions with the same parameters as that for Fig. \ref{fig:zp_insertions}. Similar to the preceding experiments, in this section, the storage size of the second version includes raw data and zero pads. For the Rsync based method, zero pads appear at the end to generate $k = 8$ number of chunks from $V = 3871$ units of data. Since, for this experiment, the total number of zero pads is held constant for the two schemes, the comparison is fair. In addition to showcasing the savings of DEC, we present in Fig. \ref{fig:rsync_vs_dec_insertions} the savings of the two-level DEC scheme where only two erasure codes are employed to cater different levels of sparsity. For such a case, the threshold $T_{\small{\rm{opt}}}$ is empirically computed based on the insertion distribution. The plots presented in Fig. \ref{fig:rsync_vs_dec_insertions} highlight the storage savings of both the $\frac{k}{2}$-level DEC and two-level DEC with respect to Rsync, against bursty insertions (with parameter $D$). However, for distributed single insertions (with parameter $P$), only the $\frac{k}{2}$-level DEC outperforms Rsync, but not the two-level DEC.

\section{Concluding remarks}
\label{sec:conc}

This paper proposes differential erasure coding techniques for improving storage efficiency and I/O reads while archiving multiple versions of data. Our evaluations demonstrate tremendous savings in storage. Moreover, in comparison to a system storing every version individually, the optimized reverse DEC retains the same I/O performance for reading the latest version (which is most typical), while reducing significantly the I/O overheads when all versions are accessed, in lieu of minor deterioration for fetching specific older versions (an infrequent event). Future works aim at integrating the proposed framework to full-fledged version management systems.

%
%

\ifCLASSOPTIONcaptionsoff
  \newpage
\fi


\begin{thebibliography}{1}
%
\bibitem{microsoft}
C. Huang, H. Simitci, Y. Xu, A. Ogus, B. Calder, P. Gopalan, J. Li, and
S. Yekhanin, ``Erasure coding in windows azure storage," in the Proc. of the \emph{USENIX Annual Technical Conference (ATC)}, 2012.
%
\bibitem{hadoop}
A. Thusoo, Z. Shao, S. Anthony, D. Borthakur, N. Jain, J. Sen Sarma,
R. Murthy, and H. Liu, ``Data warehousing and analytics infrastructure
at facebook," in the Proc. of the \emph{2010 ACM SIGMOD International
Conference on Management of data}, 2010.
%
\bibitem{google}
D. Ford, F. Labelle, F. I. Popovici, M. Stokely, V.-A. Truong, L. Barroso,
C. Grimes, and S. Quinlan, ``Availability in globally distributed storage
systems," in \emph{The 9th USENIX conference on Operating Systems Designand Implementation (OSDI)}, 2010.
%
\bibitem{DRWS}
A. G. Dimakis, K. Ramchandran, Y. Wu, and C. Suh, ``A survey on network codes for distributed storage," \emph{IEEE Proceedings}, vol. 99, pp. 476 –- 489, Mar. 2011.
%
\bibitem{OgD2}
F. Oggier and A. Datta, ``Coding Techniques for Repairability in Networked Distributed Storage Systems," Foundations and Trends in Communications and Information Theory, Now Publishers, 2013.
%

\bibitem{WaV}
Zhiying Wang and Viveck Cadambe, ``Multi-version Coding for Distributed Storage," Proc. of \emph{IEEE ISIT 2014}, Honalulu, U.S.A.

\bibitem{RVBS}
A. Rawat , S. Vishwanath , A. Bhowmick and E. Soljanin  ``Update efficient codes for distributed storage," \emph{IEEE Int. Symp. Inf. Theory,  2011}.
%
\bibitem{HPZV}
S. Han, H.-T. Pai, R. Zheng, and P. K. Varshney, ``Update-Efficient Regenerating Codes with Minimum Per-Node Storage," in Proc. of \emph{IEEE Int. Symp. Inform. Theory (ISIT) 2013}, Istanbul, 2013.
%
\bibitem{MWC}
A. Mazumdar, G. W. Wornell, and V. Chandar, ``Update efficient codes for error correction," in the Proc. of \emph{IEEE Int. Symp. Inform. Theory 2012}, pages 1558 -- 1562, Cambridge, MA, July 2012.
%
\bibitem{ECD}
K. S. Esmaili, A. Chiniah, A. Datta, ``Efficient updates in cross-object erasure-coded storage systems," \emph{IEEE International Conference on Big Data, 2013}, Oct. 2013 Silicon Valley, CA.
%
\bibitem{dedup}
V.Tarasov, A. Mudrankit, W. Buik, P. Shilane, G. Kuenning, E. Zadok, 
``Generating Realistic Datasets for Deduplication Analysis", in the Proc. of the \emph{2012 USENIX conference on Annual Technical Conference}. 
%
\bibitem{svn}
\url{http://subversion.apache.org/}
%
\bibitem{git}
\url{http://git-scm.com}
%
\bibitem{copyonwrite} 
\url{http://www.ibm.com/developerworks/tivoli/library/t-snaptsm1/index.html}

\bibitem{HOD}
J. Harshan, F. Oggier, A. Datta, ``Sparsity Exploiting Erasure Coding for Resilient Storage and Efficient I/O Access in Delta based Versioning Systems", preprint, \url{http://arxiv.org/abs/1411.4762}.

\bibitem{rsync}
\url{http://rsync.samba.org/}
%
\bibitem{compsens}
D.L. Donoho, ``Compressed Sensing", {\em IEEE Transactions on Information Theory},  Volume 52, Issue 4, April 2006. 
%
\bibitem{ZP}
F. Zhang, H. D. Pfister, ``Compressed sensing and linear codes over real numbers", 
{\em Information Theory and Applications Workshop (ITA)}, 2008.  
%



\bibitem{McS}
F. J. McWilliams, N.J.A. Sloane, ``The Theory of Error Correcting Codes," Amsterdam, The Netherlands: North Holland, 1977.

\bibitem{LaF}
J. Lacan and J. Fimes, ``A construction of matrices with no singular square submatrices," in the Proc. of \emph{International Conference on Finite Fields and Applications}, pp. 145 -- 147, 2003.


\end{thebibliography}
\end{document}